\documentclass[pra,aps,superbib,citeautoscript,twocolumn]{revtex4-1}

\usepackage[colorlinks=true,urlcolor=blue,citecolor=blue,linkcolor=blue]{hyperref}

\usepackage{amsmath,bm}
\usepackage{amstext}
\usepackage{epsfig}
\usepackage{xcolor}
\usepackage{subfig}
\usepackage{graphicx,amsmath,amssymb,tabularx}
\usepackage{multirow}
\usepackage{array}
\usepackage{dsfont}
\usepackage{caption}
\usepackage{cleveref}
\usepackage{ulem}
\usepackage[toc]{appendix}
\usepackage{physics}
\usepackage{mathrsfs}  
\usepackage{amsbsy}

\definecolor{dgreen}{rgb}{0,.5,0}
\definecolor{dred}{rgb}{.7,.0,.0}

\def\im{{\rm i}}

\DeclareMathAlphabet\mathbfcal{OMS}{cmsy}{b}{n}

\newcommand{\br}{\mathbf{r}}

\newcommand{\ie}{{\it i.e.}}

\newcommand{\be}{\begin{eqnarray}}
\newcommand{\ee}{\end{eqnarray}}
\DeclareMathAlphabet\mathbfcal{OMS}{cmsy}{b}{n}

\newcommand{\rev}[1]{{{#1}}}

\begin{document}

\title{
Exactly factorized molecular Kohn--Sham density functional theory}

\author{
Lucien Dupuy$^{a,b}$
}
\email{lucien.dupuy@unistra.fr}
\author{
Benjamin Lasorne$^c$
}
\author{
Emmanuel Fromager$^{a,b}$
}
\affiliation{\it 
~\\
$^a$Laboratoire de Chimie Quantique,
Institut de Chimie, CNRS / Universit\'{e} de Strasbourg,
4 rue Blaise Pascal, 67000 Strasbourg, France\\
}
\affiliation{$^b$University of Strasbourg Institute for Advanced Study,
5, all\'{e}e du G\'{e}n\'{e}ral Rouvillois, F-67083 Strasbourg, France\\}
\affiliation{
$^c$
ICGM, Univ Montpellier, CNRS, ENSCM, Montpellier, France
}


\begin{abstract}
Fromager and
Lasorne [{\it Electron. Struct.} {\bf 6} 025002 (2024)] have recently derived an in-principle exact Kohn--Sham
density functional theory (KS-DFT) of electrons and nuclei, where the nuclear density and the (so-called conditional) electronic density are mapped onto a fictitious electronically non-interacting KS molecule. In this work, we apply the exact factorization formalism to the molecular KS wavefunction, thus leading to disentangled (but coupled) marginal and conditional KS equations. We show that, while being equivalent to the original theory, these equations open new perspectives in the practical extension of regular (electronic) KS-DFT beyond the Born--Oppenheimer approximation. The importance and treatment of correlations induced in this context by second-order geometrical derivatives is also discussed.      
\end{abstract}

\maketitle



\section{Introduction}

When one needs to simulate properties and dynamics of molecules in their excited states while keeping the practical one-electron picture of Kohn-Sham Density Functional Theory~\cite{Hohenberg1964,Kohn1965,Burke2012perspectiveDFA,Teale2022_DFT_exchange} (KS-DFT), it is common practice to turn to its time-dependent extension~\cite{runge1984density,Casida_tddft_review_2012} (TDDFT) within either the real-time (RT) or linear-response (LR) frameworks according to context. While it indeed gives access to excited energy levels and non-adiabatic couplings~\cite{Send10_First-order,Ou15_First-order,Wang21_NAC-TDDFT} at a given nuclear geometry, it is less appreciated that it still formally relies on the Born--Oppenheimer (BO) approximation: the electronic structure of molecules is reduced to a purely electronic problem with the effect of nuclei entering as mere parameters of the BO Hamiltonian. Ignoring entirely the correlation between electrons and nuclei can lead to a spectacular breakdown of the approach, such as in the vicinity of conical intersections where two or more BO electronic states become degenerate~\cite{dom04,bae06,dom11,las11:460}. There, molecular states take on an intrinsically vibronic character, their structure unpredictable without a correlated electro-nuclear framework accounting for nuclear quantum effects. 

The breakdown of LR-TDDFT around ground state degeneracies can be seen as a consequence of this, as the single-reference KS-DFT ground electronic state is no longer a good 0$^{\rm th}$-order reference for a perturbative treatment~\cite{Casida_tddft_review_2012}. To escape the BO framework, two of the authors recently introduced an exact KS-DFT of the entire molecule~\cite{Fromager_2024_Density} with nuclear and conditional electronic densities as its basic variables. Contrary to legacy multi-component DFT approaches~\cite{Kreibich2001,Gidopoulos98,Butriy07,Kreibich08} (see also~\cite{Chakraborty2008_Development,delaLande2019_Multicomponent,Xu2023_First-Principles} and the references therein), the fictitious KS system is set to be an electronically-non-interacting molecule while keeping electro-nuclear and nuclear-nuclear interactions, thus avoiding the tedious development of additional functionals. The resulting molecular KS equation involves not only an electro-nuclear KS potential but also a nuclear-nuclear analog, both depending on nuclear and electronic densities, bringing non-adiabatic electro-nuclear correlation. Moreover, it does not require the match of additional basic variables between the KS and real system, such as electronic current and nuclear phase as in the approach of Gross and co-workers~\cite{Requist16_Exact,Li18_Density}. Our theory makes no {\it a priori} assumption about the functional form of the KS wavefunction. In our previous work, we presented the coupled beyond-BO electronic and nuclear KS equations obtained by inserting the Born--Huang (BH) expansion.

However, with the aim to cut practical computational methods out of the exact theory, it is sensible to write down the latter in such a way that it facilitates the introduction of controlled approximations. Molecular dynamics is amenable to a semi-classical treatment of nuclei (typically by means of trajectories) and electro-nuclear correlation based on their small mass ratio~\cite{Kapral_Ciccotti99}. In recent years, it has become clear that the BH expansion of the molecular wavefunction, while exact, is a difficult starting point to derive such mixed quantum-classical approximations. The Exact Factorization (XF) of the total wavefunction~\cite{Hunter1975_Conditional,Abedi10_EF,Abedi12_Correlated,Gidopoulos2014_Electronic} into a (marginal) nuclear wavefunction and (conditional) electronic wavefunction has proved enormously advantageous for mixed quantum-classical dynamics~\cite{Min15_Coupled-Trajectory,Filatov18_Direct,AMAG16,HLM18,HM22,VMM22,Villaseco22_Exact,DRM24} and continues to be exploited in new ways within the non-adiabatic context, such as non-adiabatic perturbation theory~\cite{NVPT_2015,Schild16_Electronic_Flux,Agostini_adia_EF_16,dmpv-zqdh,nnkr-phm5,tu2025nonadiabaticperturbationtheoryexact}.

The present paper introduces the XF representation of the molecular KS wavefunction and the exact coupled equations obeyed by the electronic and nuclear subsystems. The focus will be on the KS electronic equation which adopts a starkly different form than within the BH expansion, and than in other XF-based density functional formalisms~\cite{Requist16_Exact}.

The paper is organized as follows. After a brief review of both the XF formalism (in Sec.~\ref{sec:regular_XF_formalism}) and the molecular KS-DFT of Ref.~\citenum{Fromager_2024_Density} (in Sec.~\ref{sec:review_mKS-DFT}), the two approaches are combined in Sec.~\ref{sec:exact_XF_mKS-DFT}, thus leading to in-principle exact conditional and marginal KS equations. As shown in Sec.~\ref{sec:approx_1stdev}, neglecting second-order geometrical derivatives in the former equation leads to an unconventional one-electron beyond-BO KS equation, which is highly appealing for practical purposes. This ``first-order'' approximation is tested on a lattice model of a diatomic molecule in Sec.~\ref{sec:Hubbard}. As an outlook, we discuss in Sec.~\ref{sec:CI_type_approach} different strategies (that are general and applicable to {\it ab initio} systems) for recovering the missing correlation effects, namely those that are induced by second-order geometrical derivatives. Conclusions are finally given in Sec.~\ref{sec:conclusions}.

\section{Brief review of the exact factorization and molecular KS-DFT}

For the sake of simplicity and clarity, derivations will be detailed in the particular case of a molecule with a single nuclear degree of freedom (typically the bond distance $R$ in a diatom with a reduced mass denoted $M$). Obviously, the approach is general and therefore applicable to any molecule, with an arbitrary number of nuclear degrees of freedom.
In what follows, we shall assume the system of atomic units where, in particular, $\hbar=1$ and $m_{\rm e} =1$.

\subsection{Exact factorization of molecular wavefunctions}\label{sec:regular_XF_formalism}

In this section we briefly introduce the XF formalism of (stationary) molecular wavefunctions. We follow the derivation of Ref.~\citenum{Gidopoulos2014_Electronic}, starting from the ($N_{\rm e}$-electron) ground-state molecular Schr\"{o}dinger equation,
\be\label{eq:molecular_int_Schroedinger_eq}
\left[
\hat{T}_{\rm n}+\hat{H}^{\rm BO}(R)
\right]\Psi(R,r)=E\Psi(R,r),
\ee
where $R$ denotes the molecular geometry (here, the bond distance), $\hat{T}_{\rm n}$ is the nuclear kinetic energy operator (it reads $\hat{T}_{\rm n}\equiv -\frac{1}{2M}\frac{\partial^2}{\partial R^2}$ in this work, where $M$ is the reduced mass), and $r\equiv({\bf r}_1,{\bf r}_2,\ldots,{\bf r}_{N_\mathrm{e}})$ is the set of electronic coordinates. Note that $\br \equiv (x,y,z)$, in bold font, refers to a position in real space and $\br_i\equiv (x_i,y_i,z_i)$ is the position assigned to the $i$th electron. Spin degrees of freedom are taken into account implicitly. The (geometry-dependent) BO Hamiltonian,  
\be
\hat{H}^{\rm BO}(R)\equiv
\hat{T}_{\rm e}+\hat{W}_{\rm ee}+\hat{V}_{\rm
ne}(R)+V_{\rm nn}(R),
\ee
consists of the electronic kinetic energy operator ($\hat{T}_{\rm e}$), the electronic repulsion operator ($\hat{W}_{\rm ee}$), the nuclear-electron attraction potential,
\be
\hat{V}_{\rm
ne}(R)\equiv \sum_{i=1}^{N_{\mathrm{e}}}V_{\rm
ne}(R,\br_i),
\ee
and the nuclear-nuclear repulsion potential $V_{\rm nn}(R)$. We now consider the exactly factorized form~\cite{Hunter1975_Conditional,Abedi10_EF,Abedi12_Correlated,Gidopoulos2014_Electronic},
\be\label{exactly_fact_mol_wf}
\Psi(R,r)=\chi(R)\Phi_R(r)
\ee
of the ground-state molecular wavefunction, the so-called {\it conditional} electronic wavefunction $\Phi_R: r\mapsto \Phi_R(r)$ being normalized for any geometry $R$, \ie, 
\be\label{eq:normalization_cond_wf}
\langle\Phi_R\vert\Phi_R\rangle:=\int dr\,\left\vert\Phi_R(r)\right\vert^2=1,\,\forall R,
\ee
where we used the shorthand notation $\int dr\equiv \int d\br_1\ldots \int d\br_{N_{\mathrm{e}}}$. Consequently, the normalization of the full molecular wavefunction,
 \ie,
\be\label{eq:normalization_mol_wf}
\langle\Psi\vert\Psi\rangle=\int dR \int dr\; \vert\Psi(R,r)\vert^2=1,  
\ee
implies that of the (so-called {\it marginal}) nuclear wavefunction,
\be
\int dR\;\vert \chi(R)\vert^2=1.
\ee
Such a factorization is not unique but the normalization conditions reduce the gauge freedom to a mere self-compensating product (equal to one) of unimodular complex phase factors, where the phase may depend on $R$:
\rev{
\be \label{eq:gauge}
\begin{split}
\chi(R) &\rightarrow e^{-i\theta(R)} \chi(R)
\\
\Phi_{R}(r) &\rightarrow e^{i\theta(R)} \Phi_{R}(r)
\end{split}
\ee
}
As shown in Ref.~\citenum{Gidopoulos2014_Electronic} (see also Appendix~\ref{app:key_steps_XF}, where the key steps of the derivation are highlighted, for completeness), inserting the factorization of Eq.~(\ref{exactly_fact_mol_wf}) into the Schr\"{o}dinger Eq.~(\ref{eq:molecular_int_Schroedinger_eq}) leads to two disentangled but coupled equations, namely the conditional electronic one, which reads 
\be\label{eq:XF_conditional_eq}
\begin{split}
&\Bigg(\hat{H}^{\rm BO}(R)+\frac{\left(-\im\frac{\partial}{\partial
R}-A(R)\right)^2}{2M}
\\
&+\dfrac{1}{M}\left[\dfrac{\chi^*(R)}{\abs{\chi(R)}^2}\left(-\im\dfrac{\partial
\chi(R)}{\partial
R}\right)+A(R)\right]
\\
&.\left[-\im\dfrac{\partial}{\partial
R}-A(R)\right]
\Bigg)\Phi_R(r)=\mathcal{E}(R)\Phi_R(r),
\end{split}
\ee
and the marginal nuclear one,
\be\label{eq:regular_marg_nuclear_eq}
\begin{split}
\left[
\frac{\left(-\im\frac{\partial}{\partial
R}+A(R)\right)^2}{2M}+\mathcal{E}(R)
\right]\chi(R)=E\chi(R),
\end{split}
\ee
where 
\be\label{eq:pot_vec_def}
A(R)=\left\langle\Phi_R\middle \vert -\im \frac{\partial \Phi_R}{\partial
R}\right\rangle
\ee
plays the role of a vector potential component, as readily seen from Eq.~(\ref{eq:regular_marg_nuclear_eq}). The conditional electronic energy $\mathcal{E}(R)$, which is an exactification of the adiabatic BO potential energy surface (PES), reads as follows, according to Eq.~(\ref{eqapp:first_appearance_cond_elec_ener}), 
\be\label{eq:expression_conditional_ener}
\mathcal{E}(R)=\mathcal{E}^{\rm BO}(R)-\dfrac{A^2(R)}{2M}+\dfrac{1}{2M}\left\langle\dfrac{\partial
\Phi_R}{\partial R}\middle\vert\dfrac{\partial\Phi_R}{\partial R}\right\rangle,
\ee
where, in this context, the BO-like electronic energy is evaluated from the conditional wavefunction, \ie,
\be\label{eq:BO_ener_def}
\mathcal{E}^{\rm BO}(R)=\langle \Phi_R \vert \hat{H}^{\rm BO}(R)\vert
\Phi_R\rangle.
\ee
Let us stress that the expectedly dominant Hamiltonian term, $\mathcal{E}^{\rm BO}(R)$, is not to be confused with the usual BO ground-state PES, simply because $\Phi_R$ is not the BO ground-state electronic wavefunction.
Further, extra corrections are brought by the explicit account of $\partial \Phi_R/\partial R$, which turn $\mathcal{E}^{\rm BO}(R)$ into the effective PES, $\mathcal{E}(R)$, provided $A(R)$ is considered as a vector potential, duly incorporated into the effective nuclear momentum as in Eq.~(\ref{eq:regular_marg_nuclear_eq}), consistent with the gauge-theoretical framework of a quantum system subject to an external electromagnetic field.

In the rest of the paper we will focus on the electronic conditional Eq.~(\ref{eq:XF_conditional_eq}), where we assume that the marginal nuclear wavefunction $\chi(R)$ is known and that we would like to transform, ultimately, into one-electron KS-like equations. For that purpose, we consider the more compact form,
\be\label{eq:XF_conditional_eq_compact}
\begin{split}
\left[
\hat{H}^{\rm BO}(R)
+ \rev{\hat{U}^{\rm coup}_{\rm ne}}
\right]
\Phi_R(r)=\mathcal{E}(R)\Phi_R(r), 
\end{split}
\ee
where the nuclear-electron coupling operator, which involves geometrical derivatives \rev{(that is derivatives with respect to nuclear coordinates, the latter defining what is often called the molecular geometry)} up to second order, can be decomposed as follows
\be\label{eq:full_ne_coupling_pot_op}
\begin{split}
\rev{\hat{U}^{\rm coup}_{\rm ne}}
&\equiv
U^{\rm coup(0)}_{\rm ne}(R)+U^{\rm coup(1)}_{\rm ne}(R)\dfrac{\partial}{\partial R}
\\
&\quad
-\dfrac{1}{2M}\dfrac{\partial^2}{\partial R^2},
\end{split}
\ee
the zeroth and first-order potential contributions being [see Eq.~(\ref{eqapp:cond_elec_eq_developped_coupling_pot})] 
\be\label{eq:Uzero_interacting_def}
\begin{split}
U^{\rm coup(0)}_{\rm
ne}(R)&=-\dfrac{A^2(R)}{2M}+\dfrac{\im}{2M}\dfrac{\partial
A(R)}{\partial
R}
\\
&
\quad 
+
\dfrac{\im A(R)}{M}\dfrac{\chi^*(R)}{\abs{\chi(R)}^2}\dfrac{\partial
\chi(R)}{\partial
R}
\end{split}
\ee
and 
\be\label{eq:1st_order_pot_to_be_multiplied_deriv_R}
U^{\rm coup(1)}_{\rm ne}(R)=
-\dfrac{1}{M}\dfrac{\chi^*(R)}{\abs{\chi(R)}^2}\dfrac{\partial
\chi(R)}{\partial
R},
\ee
respectively.
Note that, for a given geometry $R$, $U^{\rm coup(0)}_{\rm ne}(R)$ is a functional of $\Phi_R$ and $\partial \Phi_R/\partial R$, through the vector potential $A(R)$ [see Eq.~(\ref{eq:pot_vec_def})]. Therefore, for a given marginal nuclear wavefunction $\chi$,
Eq.~(\ref{eq:XF_conditional_eq_compact}) is in principle a 
self-consistent one.\\

As readily seen from Eq.~(\ref{eq:full_ne_coupling_pot_op}), the conditional Eq.~(\ref{eq:XF_conditional_eq_compact}) is not a regular electronic equation, also because of the geometrical derivatives. As pointed out in Ref.~\citenum{li2025bornoppenheimertimedependentdensityfunctional} (see also the references therein) and further discussed in Sec.~\ref{sec:CI_type_approach}, second-order derivatives (last term on the right-hand side of Eq.~(\ref{eq:full_ne_coupling_pot_op})) prevent us from describing exactly the conditional electronic structure with a single Slater determinant, even when electrons do not interact among themselves, like in a KS molecule~\cite{Fromager_2024_Density}. 
In order to highlight these additional complications (when comparison is made with regular BO electronic structure theory) and possibly suggest how to deal with them (see Secs.~\ref{sec:approx_1stdev} and \ref{sec:CI_type_approach}), 
it is instructive, by analogy with second-order differential equations in time of classical mechanics, for example, to rewrite Eq.~(\ref{eq:XF_conditional_eq_compact}) as a {\it first-order} equation in $R$ instead. This can be achieved by using, as unknown quantity, the conditional wavefunction {\it and} its geometrical derivative, together, 
\be
\Phi_R(r)\rightarrow \left[
\begin{matrix}
\Phi_R(r)
\\
\frac{\partial \Phi_R(r)}{\partial R}
\end{matrix}
\right], 
\ee
thus leading to the equivalent matrix form of Eq.~(\ref{eq:XF_conditional_eq_compact}), 
\be\label{eq:cond_elec_eq_two_component_form}
\begin{split}
&\left[
\begin{matrix}
\hat{H}^{(0)}(R) & \quad U^{\rm coup(1)}_{\rm ne}(R)
-\frac{1}{2M}\frac{\partial}{\partial R}  
\\
\frac{\partial}{\partial R} & 0
\end{matrix}
\right]
\left[
\begin{matrix}
\Phi_R(r)
\\
\frac{\partial \Phi_R(r)}{\partial R}
\end{matrix}
\right]
\\
& 
=
\left[
\begin{matrix}
\mathcal{E}(R) & 0
\\
0& 1
\end{matrix}
\right]
\left[
\begin{matrix}
\Phi_R(r)
\\
\frac{\partial \Phi_R(r)}{\partial R}
\end{matrix}
\right]
,
\end{split}
\ee
where 
\be\label{eq:Hnot_of_R_in_matrix_def}
\hat{H}^{(0)}(R)=\hat{H}^{\rm BO}(R)+U^{\rm coup(0)}_{\rm
ne}(R).
\ee
Interestingly, the non-Hermitian structure of the latter matrix echoes that of $\hat{U}^{\rm coup}_{\rm ne}$ [see Eq.~(\ref{eq:full_ne_coupling_pot_op})] in the electronic Hilbert space or, more precisely, of its first-order geometrical derivative contribution (second term on the right-hand side of Eq.~(\ref{eq:full_ne_coupling_pot_op})). This feature, which has already been discussed in the literature~\cite{Li18_Density,li2025bornoppenheimertimedependentdensityfunctional}, becomes even more apparent when Eq.~(\ref{eq:cond_elec_eq_two_component_form}) is rewritten equivalently as follows,    
\be\label{eq:cond_elec_eq_two_component_form_equivalent_form}
\begin{split}
&\left[
\begin{matrix}
\hat{H}^{(0)}(R) & \quad \sqrt{2M} \, U^{\rm coup(1)}_{\rm ne}(R)
-\frac{1}{\sqrt{2M}}\frac{\partial}{\partial R}  
\\
-\frac{1}{\sqrt{2M}}\frac{\partial}{\partial R} & 0
\end{matrix}
\right]
\\
&\times\left[
\begin{matrix}
\Phi_R(r)
\\
\frac{1}{\sqrt{2M}}\frac{\partial \Phi_R(r)}{\partial R}
\end{matrix}
\right]
=
\left[
\begin{matrix}
\mathcal{E}(R) & 0
\\
0& -1
\end{matrix}
\right]
\left[
\begin{matrix}
\Phi_R(r)
\\
\frac{1}{\sqrt{2M}}\frac{\partial \Phi_R(r)}{\partial R}
\end{matrix}
\right]
.
\end{split}
\ee
While being completely equivalent to the original conditional Eq.~(\ref{eq:XF_conditional_eq_compact}), Eqs.~(\ref{eq:cond_elec_eq_two_component_form}) or  (\ref{eq:cond_elec_eq_two_component_form_equivalent_form}) immediately suggest that the full treatment of nonadiabatic effects from some (approximate) reference ground-state-like conditional wavefunction will involve couplings (that manifest themselves as off-diagonal elements in the conditional Hamiltonian matrix of Eq.~(\ref{eq:cond_elec_eq_two_component_form}), for example) with the complementary space of conditional excited states. This qualitative picture will be turned into practical computational schemes in Sec.~\ref{sec:CI_type_approach}, once the formalism has been merged (in the next section) with molecular KS-DFT and a reference (approximate) conditional wavefunction has been identified (See Sec.~\ref{sec:approx_1stdev}).

\subsection{Molecular KS-DFT}\label{sec:review_mKS-DFT}

We briefly review in this section some key ideas of the molecular KS-DFT that has been derived recently by some of the authors~\cite{Fromager_2024_Density}. Starting from the Rayleigh--Ritz variational principle, which brings the best approximate solution to the ground-state molecular Schr\"{o}dinger Eq.~(\ref{eq:molecular_int_Schroedinger_eq}) within an ansatz manifold, 
\be\label{eq:GS_mol_ener_general_form_wf}
E=\min_\psi\left\langle\hat{T}_{\rm n}+\hat{T}_{\rm e}+\hat{W}_{\rm ee}+\hat{V}_{\rm nn}+\hat{V}_{\rm ne}\right\rangle_{\psi},
\ee
where we use the shorthand notation $\langle\ldots\rangle_\psi=\langle\psi\vert \ldots\vert \psi\rangle$, $\psi: (R,r)\mapsto \psi(R,r)$ being a trial normalized {\it molecular} wavefunction (ansatz), 
the following (variationally exact at convergence) density-functional expression is obtained for the molecular energy, 
\be\label{eq:VP_KS_molecule}
\begin{split}
E&=\min_{\psi}
\Big\{\left\langle
\hat{T}_{\mathrm{n}}+\hat{T}_{\mathrm{e}}\right\rangle_{\psi}
\\
&
\quad+\mathcal{E}_{\rm Hxc}[\Gamma_{\psi},n_{\psi}]+\int dR\; V_{\mathrm{nn}}(R)\Gamma_{\psi}(R)
\\
&\quad+\int dR\;\Gamma_{\psi}(R)\int d{\bf r}\;V_{\mathrm{ne}}(R,{\bf r})n_{\psi}(R,\br)
\Big\},
\end{split}
\ee
where the electronic Hartree-exchange-correlation (Hxc) energy within the molecule is evaluated as a functional of both the nuclear  $\Gamma:R\mapsto \Gamma(R)$ and the {\it geometry-dependent} electronic $n: (R,{\bf r})\mapsto n(R,{\bf r})$ densities. For a given molecular wavefunction $\psi$, these densities read more explicitly as follows,    
\begin{subequations}
\begin{align}
\label{eq:nuclear_density_general_form_mol_wf}
&\Gamma_\psi(R):=\int dr\; \abs{\psi(R,r)}^2
\\
\label{eq:nuclear_dens_intdr1dr2_etc_general_wf}
&=\int d{\bf r}_1\int d{\bf r}_2\ldots \int d {\bf r}_{N_{\mathrm{e}}}\abs{\psi(R,\br_1,{\bf r}_2,\ldots,{\bf r}_{N_{\mathrm{e}}})}^2
\end{align}
\end{subequations}
and
\be\label{eq:eff_elec_dens_intdr2_etc_general_wf}
\begin{split}
n_{\psi}(R,\br):=&\dfrac{N_{\mathrm{e}}}{\Gamma_{\psi}(R)} 
\\
&
\times\int d{\bf r}_2\ldots \int d {\bf r}_{N_{\mathrm{e}}} \abs{\psi(R,\br,{\bf r}_2,\ldots,{\bf r}_{N_{\mathrm{e}}})}^2,
\end{split}
\ee
respectively, where we note that $\int dR\; \Gamma_\psi(R)=1$ and $\int d\br\;n_{\psi}(R,\br)=N_{\rm e},\,\forall R$.
By construction~\cite{Fromager_2024_Density}, the minimizing KS molecular wavefunction $\Psi^{\rm
KS}$ in Eq.~(\ref{eq:VP_KS_molecule}) reproduces the exact ground-state nuclear $\Gamma_0=\Gamma_\Psi$ and electronic $n_0=n_\Psi$ densities [see Eq.~(\ref{eq:molecular_int_Schroedinger_eq})], \ie,
\begin{subequations}
\label{eq:KS_mapping_densities}
\begin{align}
\Gamma_{\Psi^{\rm KS}}(R)&=\Gamma_0(R),
\\
n_{\Psi^{\rm KS}}(R,\br)&=n_0(R,\br).
\end{align}
\end{subequations}
Moreover, it fulfills the following self-consistent electronically non-interacting KS molecular equation,
\be\label{eq:KS_mol_eq}
\begin{split}
&\left(\hat{T}_{\mathrm{n}}+\hat{T}_{\mathrm{e}}+{V}^{\rm KS}_{\mathrm{nn}}(R)+
\hat{V}^{\rm KS}_{\mathrm{ne}}(R)\right)\Psi^{\rm KS}(R,r)
\\
&={E}^{\rm KS}\Psi^{\rm
KS}(R,r),
\end{split}
\ee
the nuclear-nuclear KS potential,
\be\label{eq:KS_NN_potential}
V^{\rm KS}_{\mathrm{nn}}(R)&=V_{\mathrm{nn}}(R)+V^{\rm
Hxc}_{\mathrm{nn}}[\Gamma_0,n_0](R),
\ee
and the nuclear-electron KS potential operator $\hat{V}^{\rm KS}_{\mathrm{ne}}(R)\equiv \sum_{i=1}^{N_{\rm e}}V^{\rm KS}_{\mathrm{ne}}(R,{\bf r}_i)$, where 
\be\label{eq:KS_Ne_potential}
V^{\rm KS}_{\mathrm{ne}}(R,{\bf r})&=V_{\mathrm{ne}}(R,{\bf r})+V^{\rm
Hxc}_{\mathrm{ne}}[\Gamma_0,n_0](R,{\bf r}),
\ee
being evaluated by applying Hxc density-functional derivative corrections to their regular analogs (the true physical nuclear-nuclear and nuclear-electron potentials):
\be\label{eq:Hxc_NN_pot}
\begin{split}
V^{\rm
Hxc}_{\mathrm{nn}}[\Gamma,n](R)&=
\dfrac{\delta \mathcal{E}_{\rm
Hxc}[\Gamma,n]}{\delta
\Gamma(R)}
\\
&\quad-\dfrac{1}{\Gamma(R)}\int d{\bf r}\;\dfrac{\delta
\mathcal{E}_{\rm
Hxc}[\Gamma,n]}{\delta n(R,{\bf r})}n(R,{\bf r})
\end{split}
\ee
and
\be\label{eq:Hxc_Ne_pot}
V^{\rm 
Hxc}_{\mathrm{ne}}[\Gamma,n](R,{\bf r})=
\dfrac{1}{\Gamma(R)}\dfrac{\delta \mathcal{E}_{\rm
Hxc}[\Gamma,n]}{\delta n(R,{\bf r})},
\ee
respectively~\cite{Fromager_2024_Density}. 
In the above construction, we have implicitly assumed that the nuclear and electronic density mappings of Eq.~(\ref{eq:KS_mapping_densities}) can be achieved onto a pure electronically non-interacting molecular wavefunction ($\Psi^{\rm KS}$), by analogy with regular electronic KS-DFT. We shall refer to this assumption as molecular non-interacting $V$-representability. While the non-interacting $v$-representability is now better understood in the context of electronic KS-DFT~\cite{GONIS20192772}, the problem has not been addressed yet in the case of molecular KS-DFT. This is left for future work.  
Note that, for practical purposes, standard (purely electronic) density-functional approximations (DFAs) can be recycled within the adiabatic approximation that was deduced from an exact molecular generalization of the adiabatic connection formalism in Ref.~\citenum{Fromager_2024_Density}.  
This showed that the molecular Hxc functional can be written as (see Sec. IV of Ref.~\citenum{Fromager_2024_Density}):
\begin{equation} \label{eq:ac_mdft}
    \mathcal{E}_{\rm
Hxc}[\Gamma,n] = \int dR\;\Gamma(R)\, \int_{0}^1 d\lambda \, \langle \hat{W}_{\rm ee} \rangle_{\phi_R^{\lambda}[\Gamma,n]},
\end{equation}
with $\lambda$ being the electron-electron interaction strength scaling parameter and $\phi_R^{\lambda}[\Gamma,n]: r\mapsto {\Psi^{\lambda}[\Gamma,n](R,r)}/{\sqrt{\Gamma(R)}}$ the effective (conditional) electronic wavefunction of the partially-interacting molecule. Assuming that, for all $R$, $\phi_R^{\lambda}[\Gamma,n](r)$ matches its local-in-$R$ approximation $\phi^{\lambda}[n_R](r)$, that is the ground partially-interacting BO electronic state reproducing the electronic density $n_R: \br\mapsto n(R,\br)$ at each nuclear geometry $R$, one can leverage the KS-DFT adiabatic connection formula,
\begin{equation} 
    E_{\rm Hxc}[n] = \int_0^1 d\lambda\, \langle \hat{W}_{\rm ee} \rangle_{\phi^{\lambda}[n]},
\end{equation}
where $E_{\rm Hxc}[n]$ is the regular (ground-state) electronic Hxc density functional of KS-DFT. Inserting it into Eq.~\eqref{eq:ac_mdft} yields
\be\label{eq:Hxc_mol_DFA}
\mathcal{E}_{\rm
Hxc}[\Gamma,n]\approx \int dR\;\Gamma(R)\,E_{\rm Hxc}[n_R].
\ee
 In this approximate picture, the electronic density $n_R$ is thought as being parameterized by the geometry $R$, hence the name ``adiabatic'' given to the approximation~\cite{Fromager_2024_Density}. Its limitations and remedies will be discussed in a forthcoming paper (see also Refs.~\citenum{Li18_Density,Wang_2025_Testing,li2025bornoppenheimertimedependentdensityfunctional}). Later in Sec.~\ref{sec:Hubbard}, the theory will be applied to a model system for which numerically exact densities can be evaluated and the corresponding KS potentials can be determined by reverse-engineering. 

\section{Applying the exact factorization to molecular KS-DFT}\label{sec:XF_mKS-DFT}

\subsection{Exact formulation}\label{sec:exact_XF_mKS-DFT}

While the molecular KS-DFT reviewed in Sec.~\ref{sec:review_mKS-DFT} offers a formal simplification of the electronic structure description within the molecule, calculating the full molecular KS wavefunction [see Eq.~(\ref{eq:KS_mol_eq})] is not as straightforward as in the BO approximation [see Eq.~(45) in Ref.~\citenum{Fromager_2024_Density} and its appendix]. Even though it can be BH-expanded (in terms of ground and excited KS-like determinants ~\cite{Fromager_2024_Density}), working with a compact electronic wavefunction, which would ideally require solving a KS-like equation for one (or a few) electronic configuration(s), is highly desirable in practice. To achieve this goal, we apply in this section the XF formalism sketched in Sec.~\ref{sec:regular_XF_formalism} to the molecular KS wavefunction, which from now on reads
\be\label{eq:XFzed_KS_mwf}
\Psi^{\rm KS}(R,r)=\chi^{\rm KS}(R)\Phi^{\rm KS}_R(r),
\ee
where 
\be\label{eq:cond_KS_wf_normalization_any_R}
\int dr\,\left\vert\Phi^{\rm KS}_R(r)\right\vert^2=1,\,\forall R.
\ee
Note that, unlike in alternative formulations of beyond-BO DFT based on XF~\cite{Requist16_Exact,li2025bornoppenheimertimedependentdensityfunctional}, the exactly-factorized molecular KS-DFT that follows reproduces, in principle exactly, the nuclear density but not the exact marginal nuclear wavefunction. In other words, $\chi^{\rm KS}(R)$, which is evaluated in the presence of non-interacting electrons, differs from the true marginal wavefunction $\chi(R)$ of Eq.~(\ref{exactly_fact_mol_wf}), even though
\be\label{eq:nuclear_dens_mapping_KS_marginal}
\left\vert\chi^{\rm KS}(R)\right\vert^2=\left\vert\chi(R)\right\vert^2=\Gamma_0(R),
\ee
according to Eqs.~(\ref{eq:normalization_cond_wf}), (\ref{eq:nuclear_density_general_form_mol_wf}), (\ref{eq:XFzed_KS_mwf}), and (\ref{eq:cond_KS_wf_normalization_any_R}).\\ 

The KS version of the conditional electronic equation (that we take under the matrix form of Eq.~(\ref{eq:cond_elec_eq_two_component_form})) is trivially obtained by replacing the BO Hamiltonian with its molecular KS (mKS) analog, \ie~[see Eq.~(\ref{eq:KS_mol_eq})],   
\be\label{eq:substitution_HBO_hmKS}
\hat{H}^{\rm BO}(R)\rightarrow \hat{h}^{\rm mKS}(R)\equiv \hat{T}_{\rm e}+\hat{V}^{\rm KS}_{\rm
ne}(R)+V^{\rm KS}_{\rm nn}(R),
\ee
and by proceeding with the substitutions
\begin{subequations}
\begin{align}
\chi(R)&\rightarrow \chi^{\rm KS}(R)
\\
\Phi_R(r)&\rightarrow \Phi^{\rm KS}_R(r)
\\
\label{eq:KS_vector_potential_exp}
A(R)&\rightarrow A^{\rm KS}(R)=-\im\left\langle\Phi^{\rm KS}_R\middle \vert \frac{\partial \Phi^{\rm KS}_R}{\partial
R}\right\rangle
\end{align}
\end{subequations}
in the evaluation of the conditional Hamiltonian, thus leading to the following modifications [see Eqs.~(\ref{eq:Uzero_interacting_def},\ref{eq:1st_order_pot_to_be_multiplied_deriv_R}), and~(\ref{eq:Hnot_of_R_in_matrix_def})],
\begin{subequations}
\begin{align}
U^{\rm coup(0)}_{\rm ne}(R)&\rightarrow u^{\rm coup(0)}_{\rm ne}(R)
\\
\label{eq:uzero_KS_func_of_Phi_and_Chi}
&\quad\quad\equiv u^{\rm coup(0)}_{\rm
ne}\left[\Phi^{\rm KS},\chi^{\rm KS}\right](R)
\\
\label{eq:uone_KS_func_of_Chi}
U^{\rm coup(1)}_{\rm ne}(R)&\rightarrow u^{\rm coup(1)}_{\rm ne}(R)\equiv u^{\rm coup(1)}_{\rm ne}\left[\chi^{\rm KS}\right](R)
\\
\label{eq:h0_def_mKS_plus_u0}
\hat{H}^{(0)}(R)
&\rightarrow \hat{h}^{(0)}(R)\equiv \hat{h}^{\rm mKS}(R)
+u^{\rm coup(0)}_{\rm ne}(R),
\end{align}
\end{subequations}
and the conditional KS equation in its final matrix form,
\be\label{eq:exact_cond_KS_eq}
\begin{split}
&\left[
\begin{matrix}
\hat{h}^{(0)}(R) & \quad u^{\rm coup(1)}_{\rm ne}(R) 
-\frac{1}{2M}\frac{\partial}{\partial R}  
\\
\frac{\partial}{\partial R} & 0
\end{matrix}
\right]
\left[
\begin{matrix}
\Phi^{\rm KS}_R(r)
\\
\frac{\partial \Phi^{\rm KS}_R(r)}{\partial R}
\end{matrix}
\right]
\\
& 
=
\left[
\begin{matrix}
\mathcal{E}^{\rm KS}(R) & 0
\\
0& 1
\end{matrix}
\right]
\left[
\begin{matrix}
\Phi^{\rm KS}_R(r)
\\
\frac{\partial \Phi^{\rm KS}_R(r)}{\partial R}
\end{matrix}
\right].
\end{split}
\ee
Eq.~(\ref{eq:exact_cond_KS_eq}), 
which reads more explicitly as follows,
\be\label{eq:KS_cond_eq_with_second_order_deriv_scalar_form}
\begin{split}
&\left[\hat{h}^{(0)}(R)+u^{\rm coup(1)}_{\rm
ne}(R)\frac{\partial}{\partial R}
-\frac{1}{2M}\frac{\partial^2}{\partial R^2}
\right]\Phi^{\rm KS}_R(r)
\\
&
=\mathcal{E}^{\rm KS}(R)\Phi^{\rm KS}_R(r),
\end{split}
\ee
is our first key result. When combined with its marginal counterpart, which reads, by analogy with Eq.~(\ref{eq:regular_marg_nuclear_eq}), 
\be\label{eq:KS_marginal_eq}
\begin{split}
&\left[
\frac{\left(-\im\frac{\partial}{\partial
R}+A^{\rm KS}(R)\right)^2}{2M}+\mathcal{E}^{\rm KS}(R)
\right]\chi^{\rm KS}(R)
\\
&=E^{\rm KS}\chi^{\rm KS}(R),
\end{split}
\ee
it is equivalent to the original mKS Eq.~(\ref{eq:KS_mol_eq}) and exact, in the sense that its solution, the conditional KS wavefunction $\Phi^{\rm KS}_R$, reproduces the true conditional electronic density, according to Eqs.~(\ref{eq:eff_elec_dens_intdr2_etc_general_wf}), (\ref{eq:KS_mapping_densities}), (\ref{eq:XFzed_KS_mwf}), and (\ref{eq:cond_KS_wf_normalization_any_R}):
\be\label{eq:elec_dens_mapping_KS_cond}
\begin{split}
n_{\Phi^{\rm KS}_R}(\br)&=
N_{\mathrm{e}}
\int d{\bf r}_2\ldots \int d {\bf r}_{N_{\mathrm{e}}} \left\vert{\Phi^{\rm KS}_R(\br,{\bf r}_2,\ldots,{\bf r}_{N_{\mathrm{e}}})}\right\vert^2
\\
&=n_{\Psi^{\rm KS}}(R,\br)=n_0(R,\br).
\end{split}
\ee

For completeness, let us note that the total energy $E^{\rm KS}$ of the fictitious KS molecule [see Eq.~(\ref{eq:KS_marginal_eq})] is {\it not} equal to that of the true molecule $E$. Indeed, according to Eqs.~(\ref{eq:VP_KS_molecule}), (\ref{eq:KS_mol_eq}), (\ref{eq:KS_NN_potential}), (\ref{eq:KS_Ne_potential}), (\ref{eq:nuclear_dens_mapping_KS_marginal}), and (\ref{eq:elec_dens_mapping_KS_cond}),
\be
\begin{split}
E&=E^{\rm KS}+\mathcal{E}_{\rm Hxc}\left[\left\vert\chi^{\rm
KS}\right\vert^2,n_{\Phi^{\rm KS}}\right]
\\
&\quad-\int dR\,
\vert\chi^{\rm KS}(R)\vert^2 V^{\rm Hxc}_{\rm nn}(R)
\\
&\quad -\int dR\,
\vert\chi^{\rm KS}(R)\vert^2\int d\br\, V^{\rm Hxc}_{\rm
ne}(R,\br)n_{\Phi^{\rm KS}_R}(\br), 
\end{split}
\ee
where the density dependence of both Hxc potentials has been dropped for compactness and $n_{\Phi^{\rm KS}}: R,\br \mapsto n_{\Phi^{\rm KS}_R}(\br)$. The same statement holds for the conditional electronic energy, whose KS analog reads [see Eqs.~(\ref{eq:expression_conditional_ener}) and (\ref{eq:substitution_HBO_hmKS})]
\be\label{eq:expression_KS_conditional_ener}
\begin{split}
\mathcal{E}^{\rm KS}(R)&=
\langle \Phi^{\rm KS}_R \vert \hat{h}^{\rm mKS}(R)\vert
\Phi^{\rm KS}_R\rangle
-\dfrac{\left(A^{\rm KS}(R)\right)^2}{2M}
\\
&\quad+\dfrac{1}{2M}\left\langle\dfrac{\partial
\Phi^{\rm KS}_R}{\partial R}\middle\vert\dfrac{\partial\Phi^{\rm KS}_R}{\partial R}\right\rangle.
\end{split}
\ee
Finally, according to the marginal KS Eq.~(\ref{eq:KS_marginal_eq}) the above energies can be related as follows 
\be
\begin{split}
E^{\rm KS}&=\mathcal{E}^{\rm KS}(R)
\\
&\quad+\dfrac{1}{\chi^{\rm KS}(R)}\frac{\left(-\im\frac{\partial}{\partial
R}+A^{\rm KS}(R)\right)^2}{2M}\chi^{\rm KS}(R),
\end{split}
\ee
where we remind the reader that $E^{\rm KS}$ is a real number.

\subsection{Approximate KS-like one-electron conditional equation}
\label{sec:approx_1stdev}
From now on we will focus on solving the conditional KS Eq.~(\ref{eq:exact_cond_KS_eq}) for a given marginal KS wavefunction $\chi^{\rm KS}$. In the light of the discussion that motivated the matrix form of the true conditional Eq.~(\ref{eq:cond_elec_eq_two_component_form}), \rev{one would be tempted to split} the conditional KS Hamiltonian matrix as follows,   
\be\label{eq:decomp_exact_cond_KS_eq_NGI}
\begin{split}
&
\left[
\begin{matrix}
\hat{h}^{(0)}(R) & \quad u^{\rm coup(1)}_{\rm ne}(R)
-\frac{1}{2M}\frac{\partial}{\partial R}  
\\
\frac{\partial}{\partial R} & 0
\end{matrix}
\right]
\\
& 
=
\left[
\begin{matrix}
\hat{h}^{(0)}(R) & \quad u^{\rm coup(1)}_{\rm ne}(R) 
\\
\frac{\partial}{\partial R} & 0
\end{matrix}
\right]
+\left[
\begin{matrix}
0 & 
-\frac{1}{2M}\frac{\partial}{\partial R}  
\\
0 & 0
\end{matrix}
\right]
,
\end{split}
\ee
thus isolating second-order geometrical derivatives in the second term on the right-hand side.
The existence of the latter term echoes the usual considerations related to the account (or not) of diagonal BO corrections (DBOCs) and reduced mass corrections within a BH picture~\cite{Rodolphe17PTmass,Matyus19DBOC,Doriol22DBOC}, which are known to be usually small (perturbative) around equilibrium geometries (far from crossings) and often neglected within the BO approximation.

\rev{However, it is desirable to operate the splitting in such a way that the gauge invariance of XF equations is separately obeyed by each term. The electro-nuclear coupling operator as a whole is gauge invariant, and it is easy to show that the usual decomposition
\be\label{eq:GI_U_decomp}
\begin{split}
\hat{u}^{\rm coup}_{\rm ne} 
= \hat{u}^{\rm coup[1]}_{\rm ne,GI} + \hat{u}^{\rm coup[2]}_{\rm ne,\rm GI}
\end{split}
\ee
with terms involving geometrical derivatives up to first order 
\be\label{eq:full_ne_coupling_pot_op_1st}
\hat{u}^{\rm coup[1]}_{\rm ne,GI}
\equiv
u^{\rm coup(1)}_{\rm ne,GI}(R)
\, .\left[-\im\dfrac{\partial}{\partial
R}- A^{\rm KS}(R)\right]
\ee
with
\be \label{eq:ne_c_pot_1st}
u^{\rm coup(1)}_{\rm ne,GI}(R) =
\dfrac{1}{M}\left[-\dfrac{\im}{\chi^{\rm KS}(R)}\dfrac{\partial
\chi^{\rm KS}(R)}{\partial
R}+A^{\rm KS}(R)\right]
\ee
and second order
\be\label{eq:full_ne_coupling_pot_op_2nd}
\hat{u}^{\rm coup[2]}_{\rm ne,GI}
\equiv
\frac{\left(-\im\frac{\partial}{\partial
R}-A^{\rm KS}(R)\right)^2}{2M}
\ee
preserves the gauge invariance for both contributions separately. On that basis, the gauge-invariant analog to Eq.~\eqref{eq:decomp_exact_cond_KS_eq_NGI} thus reads:
\be\label{eq:decomp_exact_cond_KS_eq_GI}
\begin{split}
&
\left[
\begin{matrix}
\hat{h}^{(0)}(R) & \quad u^{\rm coup(1)}_{\rm ne}(R) 
-\frac{1}{2M}\frac{\partial}{\partial R}  
\\
\frac{\partial}{\partial R} & 0
\end{matrix}
\right]
\\
& 
=
\left[
\begin{matrix}
\hat{h}^{\rm mKS}(R) - u^{\rm coup(1)}_{\rm ne,GI}(R) \,.\, A^{\rm KS}(R)  & \quad -\im \, u^{\rm coup(1)}_{\rm ne,GI}(R)
\\
\frac{\partial}{\partial R} & \quad 0
\end{matrix}
\right]
\\
& \qquad +\left[
\begin{matrix}
\frac{ \left(A^{\rm KS}(R)\right)^2}{2M}+\frac{\im}{2M}\frac{\partial A^{\rm KS}(R)}{\partial R} & \quad
\im \frac{ A^{\rm KS}(R)}{M}-\frac{1}{2M}\frac{\partial}{\partial R}  
\\
0 & \quad 0
\end{matrix}
\right]
,
\end{split}
\ee
}
While it might be treated within perturbation theory~\cite{li2025bornoppenheimertimedependentdensityfunctional,tu2025nonadiabaticperturbationtheoryexact,Schild16_Electronic_Flux}, the above matrix form indicating clearly and explicitly how the (to-be-defined) unperturbed conditional solutions would couple, a more involved diagonalization approach can also be formulated, as shown later in Sec.~\ref{sec:CI_type_approach}. In this section, we will simply neglect \rev{the second-order geometrical derivative-bearing term}, thus leading to the {\it approximate} conditional KS equation,
\rev{\be\label{eq:KS_cond_eq_no_second_order_deriv}
\begin{split}
&\left[
\begin{matrix}
\hat{h}^{\rm mKS}(R) - u^{\rm coup(1)}_{\rm ne,GI}(R) \,.\, A^{\rm KS}_{\mu}(R)  & \quad -\im \, u^{\rm coup(1)}_{\rm ne,GI}(R)
\\
\frac{\partial}{\partial R} & 0
\end{matrix}
\right]
\\
& \quad .
\left[
\begin{matrix}
\Phi^\mu_R(r)
\\
\frac{\partial \Phi^\mu_R(r)}{\partial R}
\end{matrix}
\right]
\\
& 
\quad =
\left[
\begin{matrix}
\mathcal{E}^\mu(R) & 0
\\
0& 1
\end{matrix}
\right]
\left[
\begin{matrix}
\Phi^\mu_R(r)
\\
\frac{\partial \Phi^\mu_R(r)}{\partial R}
\end{matrix}
\right]
,
\end{split}
\ee
which reads more explicitly as follows,
\be\label{eq:KS_cond_eq_no_second_order_deriv_scalar_form}
\begin{split}
\left[\hat{h}^{\rm mKS}(R) +u^{\rm coup(1)}_{\rm ne,GI}(R)
.\left(-\im\dfrac{\partial}{\partial
R}-A^{\rm KS}_{\mu}(R)\right)\right] & \Phi^\mu_R(r)
\\
=\mathcal{E}^\mu(R) \; & \Phi^\mu_R(r),
\end{split}
\ee}
where the index $\mu\geq 0$ has been introduced for labelling a specific
solution. The ground-state ($\mu=0$) solution is an approximation to the
conditional KS electronic wavefunction $\Phi^{\rm KS}_R(r)$. As
discussed in Sec.~\ref{sec:CI_type_approach}, excited-state ($\mu>0$) solutions to
this problem can be used as a basis for approaching $\Phi^{\rm
KS}_R(r)$ even further. In the context of perturbation theory, they would play the role of the perturbers.
Let us remark that the above Hamiltonian is not Hermitian, and its eigenstates are not orthogonal.

This feature can be made more explicit by solving Eq.~(\ref{eq:KS_cond_eq_no_second_order_deriv_scalar_form}) in perturbation theory, as shown in detail in Appendix~\ref{app:PT} [see, in particular, the overlap expression through first order in Eq.~(\ref{eq:non_ortho_solutions})].\\

At this point it is important to realize that solving
Eq.~(\ref{eq:KS_cond_eq_no_second_order_deriv_scalar_form}) is equivalent to solving the following one-electron KS-like equation,
\rev{\be\label{eq:one_elec_XF-based_KS_eq}
\begin{split}
&\Bigg[-\dfrac{\nabla^2_{\bf r}}{2}+V^{\rm KS}_{\rm
ne}(R,\br) \\
& + u^{\rm coup(1)}_{\rm ne,GI}(R) \,.\left(-\im\dfrac{\partial}{\partial R}-A^{(i) {\rm KS}}(R)\right)
\Bigg]\varphi^{(i)}_{R}(\br)
\\
&=\varepsilon^{(i)}(R) \,\varphi^{(i)}_{R}(\br).
\end{split}
\ee}
Indeed, any $N_{\rm e}$-electron Slater determinant
\be
\Phi^\mu_R{\equiv}\left\vert\varphi_R^{(1^\mu)}\varphi_R^{(2^\mu)}\ldots\varphi_R^{(N_{\rm e}^\mu)}\right\vert,
\ee
where the index $i^\mu\geq 1$ ($1\leq i\leq N_{\rm e}$) indicates that the (spin) orbital $\varphi_R^{(i^\mu)}$, which is taken from the complete set of solutions $\left\{\varphi_R^{(i)}: \br\mapsto \varphi_R^{(i)}(\br)\right\}_{i\geq 1}$, is occupied in $\Phi_R^\mu$, satisfies Eq.~(\ref{eq:KS_cond_eq_no_second_order_deriv_scalar_form}) with
\be\label{eq:ener_state_mu_of_R}
\mathcal{E}^\mu(R)=\sum^{N_{\rm e}}_{i=1}\varepsilon^{(i^\mu)}(R)+
V^{\rm KS}_{\rm nn}(R),
\ee
according to the definition of $\hat{h}^{\rm mKS}(R)$ [see Eq.~(\ref{eq:substitution_HBO_hmKS})]. Note that \rev{going from Eq.~\eqref{eq:KS_cond_eq_no_second_order_deriv_scalar_form} to Eq.~\eqref{eq:one_elec_XF-based_KS_eq}, the orbital-specific vector potential}
\rev{
\be \label{eq:vec_pot_orb}
A^{(i) {\rm KS}}(R) = \left\langle \varphi_R^{(i)} \middle\vert -\im \frac{\partial \varphi_R^{(i)}}{\partial
R}\right\rangle
\ee
appears upon splitting the total vector potential [see Eq.~(\ref{eq:KS_vector_potential_exp})]
\be \label{eq:KS_vector_potential_sum}
A^{KS}_{\mu}(R) = \sum_{i=1}^{N_{\rm e}} A^{(i^{\mu}) {\rm  KS}}(R)\,.
\ee}
\rev{The latter (with $\mu=0$) is still used in full in the first-order coupling term $u^{\rm coup(1)}_{\rm ne,GI}(R)$ [see Eq.~\eqref{eq:ne_c_pot_1st}]. Importantly, this ensures the approximate one-electron-like KS equation is gauge-invariant no matter how the phase is distributed between electronic orbitals.}
\rev{
\be
\begin{split}
    \Phi^\mu_{R} \rightarrow & \, e^{\im \theta(R)}  \Phi^\mu_{R} 
    \\
    & = \left\vert e^{\im \theta_{1}(R)}\varphi_R^{(1^\mu)} \, e^{\im \theta_{2}(R)}\varphi_R^{(2^\mu)}\ldots e^{\im \theta_{N_{\rm e}}(R)}\varphi_R^{(N_{\rm e}^\mu)}\right\vert
\end{split}
\ee
with $\displaystyle \sum_{i=1}^{N_{\rm e}}  \theta_i(R)  = \theta(R)$.}

\rev{As the full vector potential depends on all occupied orbitals, its presence in $u^{\rm coup(1)}_{\rm ne,GI}(R)$ might mislead the reader to think it obscures the one-electron picture we aim to achieve through Eq.~\eqref{eq:one_elec_XF-based_KS_eq}. We wish to emphasize it is no different than even the regular (BO)KS one-electron equation, which involves a KS potential depending on the electronic density, and thus on all occupied orbitals (its molecular DFT analog being $V^{\rm KS}_{\rm
ne}$).} 
We also note in passing that, despite the non-Hermiticity of the Hamiltonian in Eq.~(\ref{eq:one_elec_XF-based_KS_eq}), the $\mu$-dependent (\ie, solution-dependent) contribution to the energy (the sum on the right-hand side of Eq.~(\ref{eq:ener_state_mu_of_R})) remains real-valued \rev{as the expectation value of the first order coupling potential [second line of Eq.~(\ref{eq:one_elec_XF-based_KS_eq})] is zero by definition [see Eq.~\eqref{eq:vec_pot_orb}].}
\rev{\begin{equation}
    \left \langle \varphi^{(i)}_{R} \middle| \left(-\im\dfrac{\partial}{\partial R}-A^{(i) {\rm KS}}(R)\right)
\varphi^{(i)}_{R} \right\rangle_{\br} = 0
\end{equation}}
Eq.~(\ref{eq:one_elec_XF-based_KS_eq}) is the second key result of this work. It can be seen as a KS simplification (which is approximate for now but it can be used as reference for approaching the exact solution, see Sec.~\ref{sec:CI_type_approach}) of the true interacting many-electron conditional Eq.~(\ref{eq:XF_conditional_eq_compact}). It offers a drastically simplified
KS-like approach to non-adiabatic effects as it provides an approximation to
the beyond-BO electronic density [see Eq.~(\ref{eq:elec_dens_mapping_KS_cond})], \ie,
\be\label{eq:approx_dens_Phi_mu_equal_zero}
n_0(R,\br)\approx n_{\Phi^0_R}(\br)=\sum^{N_{\rm
e}}_{i=1}\left\vert\varphi^{(i)}_{R}(\br)\right\vert^2,
\ee
when it is solved for
the $N_{\rm e}$ lowest-in-energy
$R$-dependent (spin) orbitals $\left\{\varphi^{(i)}_{R}\right\}_{1\leq i\leq N_{\rm
e}}$. Note that, in practice, the latter should be determined self-consistently by inserting the (approximate) conditional density of Eq.~(\ref{eq:approx_dens_Phi_mu_equal_zero}) into the nuclear-electron Hxc density-functional potential [see Eq.~(\ref{eq:KS_Ne_potential})] \rev{and $u^{\rm coup(1)}_{\rm ne,GI}(R)$ be computed using Eqs.~(\ref{eq:ne_c_pot_1st},\ref{eq:vec_pot_orb},\ref{eq:KS_vector_potential_sum}) with the same set $\left\{\varphi^{(i)}_{R}\right\}_{1\leq i\leq N_{\rm
e}}$}.\\   

Let us finally stress that Eq.~(\ref{eq:one_elec_XF-based_KS_eq}) is more advanced than the beyond-BO KS Eq.~(40) of
Ref.~\citenum{Fromager_2024_Density}, because of the additional first-order geometrical derivative contribution \rev{$u^{\rm coup(1)}_{\rm ne,GI}(R) \,.\left(-\im\dfrac{\partial}{\partial R}-A^{(i) {\rm KS}}(R)\right)$}. It is also substantially different from the exactly-factorized DFT of Wang {\it
et al.}~\cite{Wang_2025_Testing} (see also Ref.~\citenum{li2025bornoppenheimertimedependentdensityfunctional}), where the beyond-BO KS potential operator remains multiplicative [see Eq.~(17) in Ref.~\citenum{Wang_2025_Testing}].

\section{Model system study}\label{sec:Hubbard}

To analyze the practicality of approximate solutions built along the lines of Sec.~\ref{sec:approx_1stdev}, we consider a model diatomic system built from the Hubbard dimer in the (diabatic) two-electron singlet basis $\Phi_1=\ket{1_{\uparrow}1_{\downarrow}}$, $\Phi_2=\frac{1}{\sqrt{2}}(\ket{1_{\uparrow}2_{\downarrow}}-\ket{1_{\downarrow}2_{\uparrow}})$, $\Phi_3=\ket{2_{\uparrow}2_{\downarrow}}$, with Hamiltonian
\begin{equation}
    \begin{bmatrix}
U-\Delta v & -\sqrt{2} \, t & 0
\\
-\sqrt{2} \, t & 0 & -\sqrt{2} \, t 
\\
0 & -\sqrt{2} \, t & U+\Delta v
\end{bmatrix}
\end{equation}
with parameters $U$, $t$, $\Delta v$ augmented with a 1D nuclear coordinate dependence interpreted as the interatomic distance $R$. We adopt the parameterization that was introduced in Ref.~\citenum{Li18_Density} but change the numerical values to bring the avoided crossing closer to $R$ regions with non-negligible nuclear density. 

\begin{figure}[h!]
    \centering
    \includegraphics[width=\linewidth]{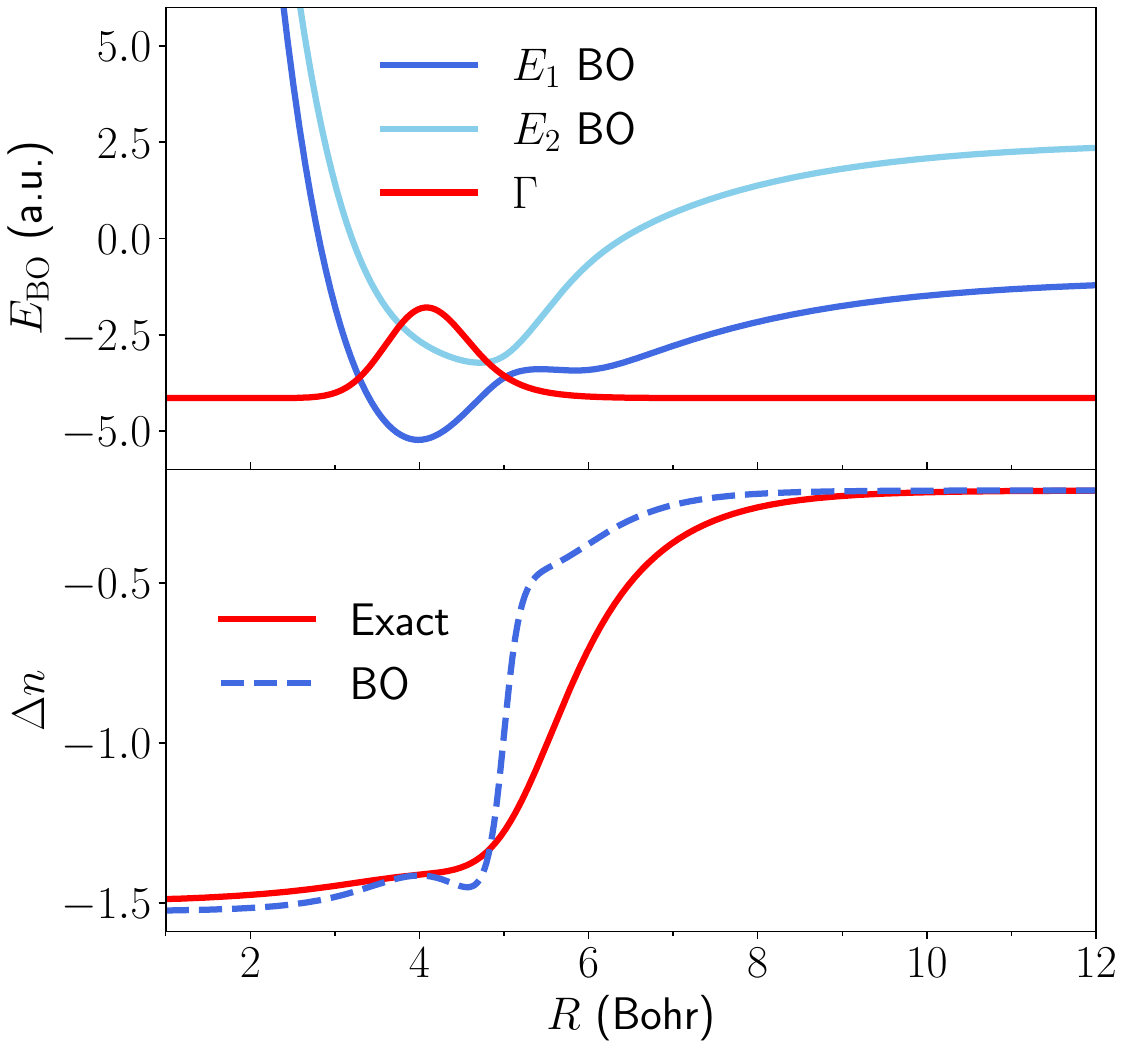}
    \caption{Top panel - BO ground (deep blue) and first excited (light blue) state PESs as a function of interatomic distance. The beyond-BO exact ground state nuclear density is shown in red. \\
    Bottom panel - BO (deep blue) and exact (red) electronic density difference between site 1 and 2 as a function of $R$.}
    \label{fig:PES_density}
\end{figure} 

The BO ground and first excited electronic PESs are drawn on top panel of Fig~\ref{fig:PES_density} together with the exact ground state nuclear density as a function of $R$. The latter was obtained by the means of a Discrete Variable Representation (DVR)~\cite{DVR_Colbert_Miller92} calculation of the molecular ground-state wavefunction. On the bottom panel, we compare the electronic density difference between atomic sites obtained in the BO approximation to the exact result. The BO solution undergoes a sharp charge transfer at the avoided crossing between $E_1$ and $E_2$, going from a strongly ionic character to an almost symmetrical split of electronic density on each atom. In contrast, the exact solution shows taking non-adiabatic effects into account displaces the distance of equilibrium between ionic and neutral character by 1 atomic unit and makes the transition much smoother. The model presents strong non-adiabatic effects while being simple enough so that we can reverse-engineer the KS potentials reproducing the exact (beyond-BO) nuclear and electronic densities. We leave the detailed exposition of our parametrization and numerical details to a follow-up publication dedicated to approximating the Kohn-Sham (KS) energy functional. Presently, we focus on developing practical approximations to the conditional KS wavefunction equation (\ref{eq:exact_cond_KS_eq}) with the nuclear-nuclear and nuclear-electron (simply referred to as electronic in the following) KS potentials, be it exact or approximate, being given. We emphasize this is an entirely separate (yet significant) matter, and we treat it as such by using the same KS potentials in the full solution to Eq.~(\ref{eq:exact_cond_KS_eq}), with a full account of second-order geometric derivatives, and its first order approximation Eq.~\eqref{eq:KS_cond_eq_no_second_order_deriv}.

\begin{figure}[h!]
    \centering
    \includegraphics[width=\linewidth]{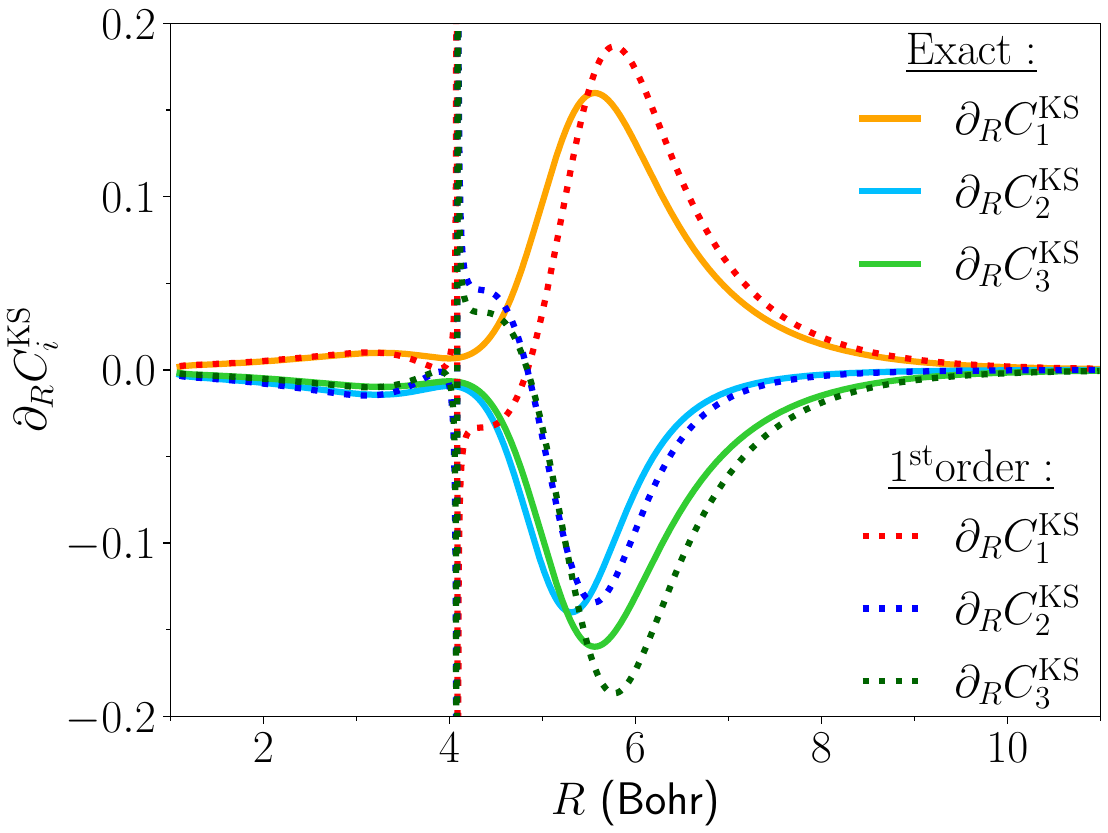}
    \caption{Comparison of the ``first-order'' approximation (dashed lines) and full resolution (full lines) of the conditional KS electronic coefficients' geometrical derivative. The corresponding key equations are Eq.~(\ref{eq:KS_cond_eq_no_second_order_deriv}), which involves first-order geometrical derivatives only and is equivalent to the one-electron conditional KS Eq.~(\ref{eq:one_elec_XF-based_KS_eq}), and the exact Eq.~\eqref{eq:exact_cond_KS_eq}, respectively.}
    \label{fig:dCKS}
\end{figure}

On Fig.~\ref{fig:dCKS} we plot the geometric first derivative of the exact KS coefficients in the singlet basis ($\displaystyle \Phi_R^{\rm KS}= \sum^3_{i=1} C^{\rm KS}_{i}(R) \;\Phi_i$). 
Those were obtained by solving the KS analog to molecular Schrödinger Eq.~\eqref{eq:molecular_int_Schroedinger_eq} in full (full lines), with exact nuclear density and electronic KS potential as input. As a comparison, we computed the prediction that Eq.~\eqref{eq:KS_cond_eq_no_second_order_deriv} [which is equivalent to the one-electron conditional KS Eq.~(\ref{eq:one_elec_XF-based_KS_eq})] gives for the coefficients' first derivative when fed with exact KS coefficients, electronic potential and nuclear density (dashed lines). 
The procedure is outlined in Appendix~\ref{sec:appHubKS}.
The singular behavior around $R=4$ a.u. corresponds to the maximum of the ground-state nuclear density and thus simply reflects that when \rev{$u^{\rm coup(1)}_{\rm ne,GI}(R)$} vanishes, Eq.~\eqref{eq:KS_cond_eq_no_second_order_deriv} does not tell us any information about the change of the electronic wavefunction at this position. There, it simplifies to a (local) eigenvalue problem with a KS electronic potential imbuing non-local (beyond-BO) effects. Thus, would the approximate solution be satisfactory everywhere else, no pathologic behavior would actually emerge. On that point, the prediction of Eq.~\eqref{eq:KS_cond_eq_no_second_order_deriv} is reasonably close to the exact one so that our proposed decomposition into a "first order" solution (Eq.~(\ref{eq:one_elec_XF-based_KS_eq})) followed by a perturbative incorporation of second-order derivative contributions holds promise.

\section{Outlook: Correlating the conditional KS wavefunction}\label{sec:CI_type_approach}

While the complete (and in-principle exact) conditional KS Eq.~\eqref{eq:exact_cond_KS_eq} can be solved easily for simple lattice models, as illustrated in Sec.~\ref{sec:Hubbard}, improving the ``first-order'' approximation of Eq.~\eqref{eq:KS_cond_eq_no_second_order_deriv} in a general and {\it ab initio} setting requires formulating a correlated method for the conditional KS wavefunction. While the derivation of a perturbation theory in the present context is left for future work, we briefly elaborate in this section on a configuration interaction (CI)-type approach.\\  

The basic idea consists in expanding the conditional KS wavefunction (and, consequently, its first-order geometrical derivative) in the basis of the ground- and excited-state Slater determinants obtained from the ``first-order'' approximation of Sec.~\ref{sec:approx_1stdev} [see the KS Eq.~(\ref{eq:one_elec_XF-based_KS_eq}) and Eq.~(\ref{eq:approx_dens_Phi_mu_equal_zero})\rev{, recalling that $V^{\rm KS}_{\rm
ne}$ and $u^{\rm coup(1)}_{\rm ne,GI}$ are to be computed from the $\mu=0$ solution}],     
\be\label{eq:CI_exp_KS_cond_wf}
\left[
\begin{matrix}
\Phi^{\rm KS}_R(r)
\\
\frac{\partial \Phi^{\rm KS}_R(r)}{\partial R}
\end{matrix}
\right]=\sum_{\mu\geq 0}C_\mu\left[
\begin{matrix}
\Phi^\mu_R(r)
\\
\frac{\partial \Phi^\mu_R(r)}{\partial R}
\end{matrix}
\right]
\ee
where, for simplicity, we assume $R$-independent CI coefficients ${\bf C}=\left\{C_\mu\right\}_{\mu\geq 0}$, so that we can use the same expansion for the conditional KS wavefunction and its derivative with respect to $R$ at the present stage.
According to Eqs.~\eqref{eq:exact_cond_KS_eq} and (\ref{eq:decomp_exact_cond_KS_eq_GI},\ref{eq:KS_cond_eq_no_second_order_deriv}), they  satisfy 
\be\label{eq:cond_KS_eq_CI_form_explicit}
\begin{split}
&\sum_{\mu\geq 0}C_\mu \left[
\begin{matrix}
\rev{\bar{\mathcal{E}}^\mu(R)} & \quad
\rev{\im \frac{ A^{\rm KS}(R)}{M}}-\frac{1}{2M}\frac{\partial}{\partial R}  
\\
0& \quad 1
\end{matrix}
\right].
\left[
\begin{matrix}
\Phi^\mu_R(r)
\\
\frac{\partial \Phi^\mu_R(r)}{\partial R}
\end{matrix}
\right]
\\
&=\left[
\begin{matrix}
\mathcal{E}^{\rm KS}(R) & 0
\\ 
0& 1
\end{matrix}
\right]\sum_{\mu\geq 0}C_\mu \left[
\begin{matrix}
\Phi^\mu_R(r)
\\
\frac{\partial \Phi^\mu_R(r)}{\partial R}
\end{matrix}
\right]
,\,\forall R,\,\forall r,
\end{split}
\ee
\rev{with
\be \label{eq:KS_e_mu_CI}
\begin{split}
    \bar{\mathcal{E}}^\mu(R) = & \, \mathcal{E}^\mu(R) +\frac{ \left(A^{\rm KS}(R)\right)^2}{2M}+\frac{\im}{2M}\frac{\partial A^{\rm KS}(R)}{\partial R}
    \\
    & + u^{\rm coup(1)}_{\rm ne,GI}(R)\,.\left(A^{\rm KS}_{\mu}(R)-A^{{\rm KS}}(R)\right)
\end{split}
\ee
}
\rev{and the full-solution vector potential given by}
\rev{\be \label{eq:KS_vec_pot_CI}
    A^{\rm KS}(R)= \sum_{\mu\geq 0,\nu\geq 0}C^*_\mu C_\nu\langle \Phi^\mu_R \vert -\im \frac{\partial  \Phi^\nu_R}{\partial R}\rangle
\ee}

or, equivalently,
\be
\begin{split}
&
\sum_{\mu\geq 0}C_\mu
\left[
\begin{matrix}
\rev{\bar{\mathcal{E}}^\mu(R)} & \quad \rev{\im \frac{ A^{\rm KS}(R)}{M}}-\frac{1}{2M}\frac{\partial}{\partial R} 
\end{matrix}
\right]
\left[
\begin{matrix}
\Phi^\mu_R(r)
\\
\frac{\partial \Phi^\mu_R(r)}{\partial R}
\end{matrix}
\right]
\\
&=
\mathcal{E}^{\rm KS}(R)
\sum_{\mu\geq 0}C_\mu \Phi^\mu_R(r), \,\forall R,\,\forall r. 
\end{split}
\ee
When projected onto a given conditional KS state $\nu\geq 0$, the above equation can be written in the more compact form 
\be\label{eq:R_dependent_CI_eq}
{\bf H}^{\rm KS}(R){\bf C}=\mathcal{E}^{\rm KS}(R){\bf S}(R){\bf C}, \,\forall R,
\ee
where the geometry-dependent KS conditional Hamiltonian and overlap matrix elements read
\be\label{eq:Hks_nu_mu_of_R}
\begin{split}
&[{\bf H}^{\rm KS}(R)]_{\nu\mu}
\\
&=
\int dr\left[
\Phi^\nu_R(r)
\right]^*
\left[
\begin{matrix}
\rev{\bar{\mathcal{E}}^\mu(R)} & \quad \rev{\im \frac{ A^{\rm KS}(R)}{M}}-\frac{1}{2M}\frac{\partial}{\partial R}
\end{matrix}
\right]
\left[
\begin{matrix}
\Phi^\mu_R(r)
\\
\frac{\partial \Phi^\mu_R(r)}{\partial R}
\end{matrix}
\right]
\end{split}
\ee
and
\be
[{\bf S}(R)]_{\nu\mu}=\int dr\,\left[\Phi^\nu_R(r)\right]^*\Phi^\mu_R(r)
=\langle \Phi^\nu_R\vert \Phi^\mu_R\rangle 
,
\ee
respectively.
The non-orthogonality of the Slater determinants stems from the non-Hermitian nature of the ``first-order'' approximate Hamiltonian in Eq.~\eqref{eq:KS_cond_eq_no_second_order_deriv_scalar_form}. 
As mentioned previously, this can be made more explicit when treating the deviation from Hermiticity in perturbation theory (see Appendix~\ref{app:PT} for further details). In fact, as readily seen from Eq.~(\ref{eq:non_ortho_solutions}), the overlap matrix elements are (through first order in perturbation theory) proportional to the non-adiabatic couplings between the unperturbed solutions, \ie, the solutions to Eq.~\eqref{eq:KS_cond_eq_no_second_order_deriv_scalar_form} where the first-order coupling term \rev{$u^{\rm coup(1)}_{\rm ne,GI}(R)$} has been neglected.

In general, a non-Hermitian operator has complex eigenvalues, but cases where they happen to be real exist~\cite{Surjan24_realeignH}, as is the case of our first order approximation and of the full XF problem, due to the nature of the terms involved in assembling their expressions.\\

A working equation can finally be obtained through multiplication by the marginal nuclear density and integration over the geometry, thus leading to  
\be\label{eq:non_ortho_eigenvalue_pb_matrix_form}
{\bf H}^{\rm KS}{\bf C}=\mathcal{E}^{\rm
KS}{\bf S}{\bf C},
\ee
where the geometrically-averaged conditional KS Hamiltonian and overlap matrices are defined as follows, 
\be
{\bf H}^{\rm KS}=\int dR\,\left\vert\chi^{\rm KS}(R)\right\vert^2 {\bf H}^{\rm KS}(R)
\ee
and
\be\label{eq:geo_averaged_overlap_def}
{\bf S}=\dfrac{\int dR\,\left\vert\chi^{\rm
KS}(R)\right\vert^2\mathcal{E}^{\rm KS}(R){\bf S}(R)
}
{\int dR\,\left\vert\chi^{\rm 
KS}(R)\right\vert^2\mathcal{E}^{\rm KS}(R)},
\ee
respectively,
and $\mathcal{E}^{\rm
KS}=\int dR\,\left\vert\chi^{\rm KS}(R)\right\vert^2\mathcal{E}^{\rm KS}(R)$. Note that, according to the CI expansion of Eq.~(\ref{eq:CI_exp_KS_cond_wf}), 
\be
\begin{split}
{\bf C}^\dagger{\bf S}(R){\bf C}
&=\sum_{\nu\geq 0}\sum_{\mu\geq 0}C_\nu^*[{\bf S}(R)]_{\nu\mu}C_\mu
\\
&=\langle \Phi^{\rm KS}_R\vert\Phi^{\rm KS}_R \rangle
\\
&=1,\,\forall R,
\end{split}
\ee
so that the conditional KS PES can be determined as follows [see Eq.~(\ref{eq:R_dependent_CI_eq})],
\be\label{eq:R_dependent_cond_KS_ener_exp_value_matrix_form}
\mathcal{E}^{\rm KS}(R)\equiv \mathcal{E}^{\rm KS}[{\bf C}](R)={\bf
C}^\dagger{\bf H}^{\rm KS}(R){\bf C}.
\ee
Eq.~(\ref{eq:non_ortho_eigenvalue_pb_matrix_form}), combined with the above equation, is the third key result of the present work, whose practical implementation is left for future work. Note that the corresponding non-orthogonal CI problem is self-consistent in several ways. First of all, a trial conditional KS PES is requested in order to evaluate the overlap matrix ${\bf S}$, according to Eq.~(\ref{eq:geo_averaged_overlap_def}). For that purpose, the pure conditional ground-state determinant $\Phi^{\mu=0}_R$ can be used as a guess in Eq.~(\ref{eq:R_dependent_cond_KS_ener_exp_value_matrix_form}), \ie, ${\bf C}\rightarrow \left\{\delta_{\mu0}\right\}_{\mu\geq 0}$, thus initializing a first self-consistency loop in which ${\bf C}$ describes a correlated conditional KS state through Eq.~(\ref{eq:non_ortho_eigenvalue_pb_matrix_form}). Secondly, we note from Eqs.~(\ref{eq:ener_state_mu_of_R}) and (\ref{eq:Hks_nu_mu_of_R}) that ${\bf C}$ should also be updated in the calculation of the geometry-dependent conditional KS Hamiltonian matrix ${\bf H}^{\rm KS}(R)$, through the evaluation of the KS vector potential \rev{[see Eqs.~(\ref{eq:KS_e_mu_CI},\ref{eq:KS_vec_pot_CI},\ref{eq:Hks_nu_mu_of_R})]} {\it and} the conditional density that is inserted into the nuclear-nuclear KS density-functional potential [see Eqs.~(\ref{eq:KS_NN_potential}) and (\ref{eq:elec_dens_mapping_KS_cond})]. Finally, updating the conditional density in the nuclear-electron KS density-functional potential [see Eq.~(\ref{eq:KS_Ne_potential})] \rev{and the full-solution KS vector potential in the first-order coupling term [see Eq.~(\ref{eq:ne_c_pot_1st})]} will have an impact on the orbitals (which could still be determined self-consistently {\it via} the correlated conditional density expression) and their energies, according to Eq.~(\ref{eq:one_elec_XF-based_KS_eq}).\\

As a final note, let us point out that a possibly more accurate one-electron orbital approximation to Eq.~\eqref{eq:KS_cond_eq_with_second_order_deriv_scalar_form} would consist in only neglecting second-order derivative terms involving two different orbitals of the conditional wavefunction. Indeed, by writing
\begin{equation} \label{eq:sepsecdev}
    \begin{split}
            \frac{\partial^2 \Phi^{\mu}_R}{\partial R^2} = &  \sum^{N_{\rm
e}}_{i=1}\left\vert\varphi_R^{(1^\mu)}\ldots\dfrac{\partial^2
\varphi_R^{(i^\mu)}}{\partial
R^2}\ldots \varphi_R^{(N_{\rm e}^\mu)}\right\vert
\\
 & + \sum_{i \neq j}^{N_e} \left\vert\varphi_R^{(1^\mu)}\ldots\dfrac{\partial
\varphi_R^{(i^\mu)}}{\partial
R} \ldots\dfrac{\partial
\varphi_R^{(j^\mu)}}{\partial
R}\ldots \varphi_R^{(N_{\rm e}^\mu)}\right\vert
,
    \end{split}
\end{equation}
we see that neglecting the cross derivative terms (on the second line of the above equation) in the evaluation of second-order derivatives in Eq.~\eqref{eq:KS_cond_eq_with_second_order_deriv_scalar_form} \rev{ would allow us to recover a one-electron-like picture.} 
This motivates the separation of second-order geometrical derivative terms into a one-electron component (first line of Eq.~\eqref{eq:sepsecdev}),
\be
\frac{\partial^{(1)}}{\partial
R}
\dfrac{\partial \tilde{\Phi}^\mu_R}{\partial R}
\equiv\sum^{N_{\rm
e}}_{i=1}\left\vert\tilde{\varphi}_R^{(1^\mu)}\ldots\dfrac{\partial^2
\tilde{\varphi}_R^{(i^\mu)}}{\partial
R^2}\ldots \tilde{\varphi}_R^{(N_{\rm e}^\mu)}\right\vert
,
\ee
and the cross-derivative terms as the remainder,
\be\label{eq:remainder_cross_deriv}
\frac{\partial^{(2)}}{\partial
R}\frac{\partial \tilde{\Phi}^\mu_R}{\partial R}=
\dfrac{\partial^2\tilde{\Phi}^\mu_R}{\partial R^2}
-\frac{\partial^{(1)}}{\partial
R}
\dfrac{\partial \tilde{\Phi}^\mu_R}{\partial R}
.
\ee
The above contribution describes the first-order geometrical derivative of two different orbitals within the Slater determinant $\tilde{\Phi}^\mu_R{\equiv}
\left\vert\tilde{\varphi}_R^{(1^\mu)}\tilde{\varphi}_R^{(2^\mu)}\ldots\tilde{\varphi}_R^{(N_{\rm e}^\mu)}\right\vert$ and does model, as such, correlation effects. \rev{To perform this approximation in a way that preserves the gauge invariance of the resulting equation, we split $\hat{u}^{\rm coup[2]}_{\rm ne,GI}$ as
\be \label{eq:u2GI_split}
   \hat{u}^{\rm coup[2]}_{\rm ne,GI} =  \sum_{i=1}^{N_{\rm e}} \frac{\left(-\im \frac{\partial^{(i)}}{\partial R} -A^{(i) {\rm  KS}}(R)\right)^2}{2M} + \hat{u}^{\rm coup[2]}_{\rm nec,GI}
\ee
where it is understood that 
\be 
\frac{\partial^{(i)}}{\partial R} \tilde{\Phi}^{\mu}_R = \sum_{i=1}^{N_{\rm e}} \left\vert\tilde{\varphi}_R^{(1^\mu)}\ldots\dfrac{\partial
\tilde{\varphi}_R^{(i^\mu)}}{\partial
R}\ldots \tilde{\varphi}_R^{(N_{\rm e}^\mu)}\right\vert
\ee
and $\frac{\partial^{(i)}}{\partial R}$ differentiates $A^{(i) {\rm  KS}}(R)$ in the same way as $\frac{\partial}{\partial R}$. The nuclei-mediated two-electrons correlation part reads
\be
\begin{split}
    \hat{u}^{\rm coup[2]}_{\rm nec,GI} = & \frac{1}{2M} \sum_{i \neq j}^{N_{\rm e}} \left( A^{(i) {\rm  KS}}(R) A^{(j) {\rm  KS}}(R) - \frac{\partial^{(i)}\partial^{(j)}}{\partial R^2} \right) \\
    & + \frac{\im}{M} \sum_{i \neq j}^{N_{\rm e}} A^{(i) {\rm  KS}}(R) \frac{\partial^{(j)}}{\partial R} \,.
\end{split}
\ee
Both terms in Eq.~\eqref{eq:u2GI_split} are gauge-invariant for arbitrary distribution of the phase between orbitals.
Introducing $A^{(1) {\rm  KS}}(R)$ such that
\be
\begin{split}
    A^{(1) {\rm  KS}}(R) & \dfrac{\partial \tilde{\Phi}_R}{\partial R}
    \\
    & \equiv\sum^{N_{\rm e}}_{i=1} A^{(i) {\rm  KS}}(R) . \left\vert\tilde{\varphi}_R^{(1)}\ldots\dfrac{\partial
    \tilde{\varphi}_R^{(i)}}{\partial R}\ldots \tilde{\varphi}_R^{(N_{\rm e})}\right\vert ,
\end{split}
\ee
with the ``cross term'' remainder
\be
A^{(2) {\rm  KS}}(R)  \dfrac{\partial \tilde{\Phi}_R}{\partial R}=
A^{\rm  KS}(R) \dfrac{\partial \tilde{\Phi}_R}{\partial R}
-A^{(1) {\rm  KS}}(R)
\dfrac{\partial \tilde{\Phi}_R}{\partial R}
,
\ee
the gauge-invariant one/two-electron(s) splitting of $\hat{u}^{\rm coup[2]}_{\rm ne,GI}$ in matrix form reads
\be \label{eq:u2split}
\begin{split}
   & \left[
    \begin{matrix}
        \frac{ \left(A^{\rm KS}(R)\right)^2}{2M} + \frac{\im}{2M}\frac{\partial A^{\rm KS}(R)}{\partial R} 
        & \quad  {\scriptstyle \im} \frac{ A^{\rm KS}(R)}{M}-\frac{1}{2M}\frac{\partial}{\partial R}  
        \\
        0 & \quad 0
    \end{matrix}
    \right] 
\\
    & = \left[
    \begin{matrix}
        {\displaystyle \sum_{i}^{N_{\rm e}}} \frac{\left(A^{(i) {\rm  KS}}(R)\right)^2}{2M} + \frac{\im \partial_R A^{\rm KS}(R)}{2M}
        & \quad  {\scriptstyle \im}\frac{ A^{(1) {\rm  KS}}(R) }{M} -\frac{1}{2M}\frac{\partial^{(1)}}{\partial R}
        \\
        0 & \quad 0
    \end{matrix}
    \right]
    \\
    & \quad + \left[
    \begin{matrix}
        {\displaystyle \sum_{i \neq j}^{N_{\rm e}}} \frac{A^{(i) {\rm  KS}}(R) A^{(j){\rm  KS}}(R)}{2M} 
        & \quad {\scriptstyle \im} \frac{A^{(2) {\rm  KS}}(R)}{M} - \frac{1}{2M}\frac{\partial^{(2)}}{\partial R}
        \\
        0 & \quad 0
    \end{matrix}
    \right]
\end{split}
\ee
We obtain the approximate one-electron KS-like equation
\be\label{eq:partial_second_order_approx_1elec}
\begin{split}
& \Bigg[-\dfrac{\nabla^2_{\bf r}}{2}+V^{\rm KS}_{\rm
ne}(R,\br) + \frac{\left(-\im \frac{\partial^{(i)}}{\partial R} -A^{(i) {\rm  KS}}(R)\right)^2}{2M} \\
& \qquad + u^{\rm coup(1)}_{\rm ne,GI}(R)\left(-\im\dfrac{\partial}{\partial R}-A^{(i) {\rm KS}}(R)\right)\Bigg]\tilde{\varphi}^{(i)}_{R}(\br)
\\
& \quad =\tilde{\varepsilon}^{(i)}(R)\tilde{\varphi}^{(i)}_{R}(\br),
\end{split}
\ee
\rev{with approximate total KS energy
\be
    \tilde{\mathcal{E}}^\mu(R) = & \sum^{N_{\rm e}}_{i=1}\tilde{\varepsilon}^{(i^\mu)}(R)+
V^{\rm KS}_{\rm nn}(R),
\ee }
which is expected to be more accurate than Eq.~\eqref{eq:KS_cond_eq_no_second_order_deriv}, while still being amenable to a one-electron computational treatment.}

\rev{The KS orbital energies $\tilde{\varepsilon}^{(i^\mu)}(R)$ are (as they should) real no matter the gauge. Indeed, we recall $\left\langle \tilde{\varphi}^{(i)}_{R} \middle| u^{\rm coup(1)}_{\rm ne,GI}(R)\left(-\im\dfrac{\partial}{\partial R}-A^{(i) {\rm KS}}(R)\right) \tilde{\varphi}^{(i)}_{R} \right\rangle_{\br} =0$ by definition, and the expectation value of the approximate second order term reads}
\rev{
    $\frac{1}{2M} \left[ \left\langle \frac{\partial \tilde{\varphi}^{(i)}_{R}}{\partial R} \middle| \frac{\partial \tilde{\varphi}^{(i)}_{R}}{\partial R} \right\rangle_{\br} -\left(A^{(i) {\rm  KS}}(R)\right)^2 \right]$
}
\rev{which is real.}
Finally comes the question of how to generate the in-principle exact correlated conditional KS wavefunction $\Phi^{\rm KS}_R$ from the alternative basis of Slater determinants $\{\tilde{\Phi}^\mu_R\}$. In fact, we can simply proceed by analogy with Eq.~(\ref{eq:cond_KS_eq_CI_form_explicit}). Indeed, from the following expansion, 
\be
\left[
\begin{matrix}
\Phi^{\rm KS}_R(r)
\\
\frac{\partial \Phi^{\rm KS}_R(r)}{\partial R}
\end{matrix}
\right]=\sum_{\mu\geq 0}\tilde{C}_\mu\left[
\begin{matrix}
\tilde{\Phi}^\mu_R(r)
\\
\frac{\partial \tilde{\Phi}^\mu_R(r)}{\partial R}
\end{matrix}
\right]
,
\ee
and Eq.~(\ref{eq:partial_second_order_approx_1elec}), the exact conditional KS Eq.~(\ref{eq:exact_cond_KS_eq}) becomes, according to Eq.~(\ref{eq:u2split}),
\rev{\be\label{eq:alternative_decomp_exact_cond_KS_eq}
\begin{split}
&\sum_{\mu\geq 0}\tilde{C}_\mu \Bigg( \left[
\begin{matrix}
\tilde{\mathcal{E}}^\mu(R) & \quad 0
\\
0 & \quad 1
\end{matrix}
\right] 
\\
& + \left[
    \begin{matrix}
        {\scriptstyle \left(u^{\rm coup(1)}_{\rm ne,GI}(R) - \frac{\im \partial_R}{2M}  \right) \, \Delta A^{\rm KS}_{\mu}(R)} - \frac{\Delta A^{\rm KS}_{\mu,2}(R)}{2M} & \quad -\frac{\im}{M} {\scriptstyle \Delta A^{\rm KS}_{\mu}(R) }
        \\
        0 & \quad 0
    \end{matrix}
    \right]
\\
& + \left[
    \begin{matrix}
        {\displaystyle \sum_{i \neq j}^{N_{\rm e}}} \frac{A^{(i)(j) {\rm  KS}}(R)}{2M} 
        & \quad {\scriptstyle \im} \frac{A^{(2) {\rm  KS}}(R)}{M} - \frac{1}{2M}\frac{\partial^{(2)}}{\partial R}
        \\
        0 & \quad 0
    \end{matrix}
    \right] \Bigg) .
\left[
\begin{matrix}
\tilde{\Phi}^\mu_R(r)
\\
\frac{\partial \tilde{\Phi}^\mu_R(r)}{\partial R}
\end{matrix}
\right]
\\
& = \left[
\begin{matrix}
\mathcal{E}^{\rm KS}(R) & 0
\\ 
0& 1
\end{matrix}
\right]\sum_{\mu\geq 0}\tilde{C}_\mu \left[
\begin{matrix}
\tilde{\Phi}^\mu_R(r)
\\
\frac{\partial \tilde{\Phi}^\mu_R(r)}{\partial R}
\end{matrix}
\right]
,\,\forall R,\,\forall r,
\end{split}
\ee}
\rev{with shorthand notations $ \Delta A^{\rm KS}_{\mu}(R) := A^{\rm KS}_{\mu}(R) - A^{\rm KS}(R) $, $\Delta A^{\rm KS}_{\mu,2}(R) := \left(A^{\rm KS}_{\mu}(R)\right)^2 - \left(A^{\rm KS}(R)\right)^2$ and $A^{(i)(j) {\rm  KS}}(R) := A^{(j) {\rm  KS}}(R) A^{(j) {\rm  KS}}(R)$.
}
Equivalently, \rev{we can write it as}
\begin{widetext}
\rev{\be
\sum_{\mu\geq 0}\tilde{C}_\mu\left[
\begin{matrix}
\Delta\tilde{\mathcal{E}}^\mu(R)  +  {\displaystyle \sum_{i \neq j}^{N_{\rm e}}} \frac{A^{(i)(j) {\rm  KS}}(R)}{2M}  & \quad {\scriptstyle \im} \frac{A^{(2) {\rm  KS}}-\Delta A^{\rm  KS}_\mu(R)}{M} -\frac{1}{2M}\frac{\partial^{(2)}}{\partial
R} 
\end{matrix}
\right]
\left[
\begin{matrix}
\tilde{\Phi}^\mu_R(r)
\\
\frac{\partial \tilde{\Phi}^\mu_R(r)}{\partial R}
\end{matrix}
\right]
=
\mathcal{E}^{\rm KS}(R)\sum_{\mu\geq 0}\tilde{C}_\mu
\tilde{\Phi}^\mu_R(r)
,\,\forall R,\,\forall r,
\ee}
\end{widetext}
\rev{with}
\rev{\begin{equation}
    \begin{split}
            \Delta \tilde{\mathcal{E}}^\mu(R) = & \, \tilde{\mathcal{E}}^\mu(R) + \left(u^{\rm coup(1)}_{\rm ne,GI}(R) - \frac{\im \partial_R}{2M}  \right) \, \Delta A^{\rm KS}_{\mu}(R)
            \\
            & - \frac{\Delta A^{\rm KS}_{\mu,2}(R)}{2M}
    \end{split}
\end{equation}}
which can be solved along the same lines as before [see Eqs.~(\ref{eq:R_dependent_CI_eq})-(\ref{eq:R_dependent_cond_KS_ener_exp_value_matrix_form})].

\section{Conclusions}\label{sec:conclusions}

An exactly-factorized formulation of the molecular KS-DFT recently proposed by two of the authors~\cite{Fromager_2024_Density} has been derived. While being equivalent (and in-principle exact) to the original electronically non-interacting molecular KS Eq.~(\ref{eq:KS_mol_eq}), the resulting conditional and marginal KS equations [Eqs.~(\ref{eq:KS_cond_eq_with_second_order_deriv_scalar_form}) and~(\ref{eq:KS_marginal_eq}), respectively] open new perspectives in the extension of regular (electronic) DFT beyond the BO approximation. In particular, when second-order geometrical derivatives are neglected, the conditional non-interacting many-electron problem can be recast exactly into the one-electron KS-like Eq.~(\ref{eq:one_elec_XF-based_KS_eq}), where the beyond-BO KS potential operator, which is usually multiplicative~\cite{Wang_2025_Testing}, now incorporates first-order geometrical derivatives. The reasonably good performance of this ``first-order'' approximation on a lattice model for diatomics suggests that the missing correlation effects, which are induced by the second-order geometrical derivatives, could be retrieved from perturbation theory. A (possibly more robust and general) non-orthogonal CI treatment is also possible, as detailed in the previous section, for completeness [see Eq.~(\ref{eq:non_ortho_eigenvalue_pb_matrix_form})]. While it is natural to use the ``first-order'' approximation mentioned above as reference, for generating the (non-orthogonal) set of orbitals, second-order geometrical derivatives might be (partially) incorporated into a new set of KS-like orbitals, as shown in Eq.~(\ref{eq:partial_second_order_approx_1elec}). A perturbative treatment of correlation effects, or a truncated CI expansion, where a few (ground and low-lying) electronic configurations are taken into account, might be even more relevant in this case. Finally, in the light of the very recent Ref.~\citenum{li2025bornoppenheimertimedependentdensityfunctional}, the theory should be extended to the time-dependent regime, thus broadening its applicability to non-adiabatic dynamics simulations, for example. Work in these directions is currently in progress.

\section*{Acknowledgements}

This work has benefited from support provided by the University of Strasbourg Institute for Advanced Study (USIAS) for a Fellowship, within the French national programme “Investment for the future” (IdEx-Unistra).

\section*{Conflicts of Interest}

The authors declare no conflicts of interest.

\section*{Data Availability}
The data that support the findings of this study are available from the corresponding author upon reasonable request.

\appendix

\section{Key steps in the derivation of the conditional and marginal equations}\label{app:key_steps_XF}

Starting from the molecular Schrödinger Eq.~(\ref{eq:molecular_int_Schroedinger_eq}), we obtain the following nuclear equation, after multiplying by $\Phi^*_R(r)$ and integrating over the electronic coordinates $r$ [see Eqs.~(\ref{eq:normalization_cond_wf}), (\ref{eq:pot_vec_def}) and (\ref{eq:BO_ener_def})], 
\be
\begin{split}
&-\dfrac{1}{2M}\int dr\,\Phi^*_R(r)\dfrac{\partial^2 \left[\chi(R)\Phi_R(r)\right]}{\partial R^2}
+\mathcal{E}^{\rm BO}(R)\chi(R)
\\
&=E\chi(R),
\end{split}
\ee
or, equivalently,
\be\label{eqapp:developed_nuclear_eq}
\begin{split}
&-\dfrac{1}{2M}\dfrac{\partial^2 \chi(R)}{\partial R^2}
-\dfrac{\im}{M}A(R)\dfrac{\partial \chi(R)}{\partial R}
\\
&+
\left(
\mathcal{E}^{\rm BO}(R)
-\dfrac{1}{2M}\left\langle \Phi_R\middle\vert
\dfrac{\partial^2 \Phi_R}{\partial R^2}\right\rangle
\right)\chi(R)
=E\chi(R).
\end{split}
\ee
Since
\be\label{eqapp:expansion_momentum_plus_A_square}
\begin{split}
&\frac{\left(-\im\frac{\partial}{\partial
R}+A(R)\right)^2}{2M}
\\
&=
\frac{
1}{2M}
\left(-\im\frac{\partial}{\partial
R}+A(R)\right)
\left(-\im\frac{\partial}{\partial
R}+A(R)\right)
\\
&=\dfrac{A^2(R)}{2M}-\dfrac{\im}{2M} \left(\dfrac{\partial
A(R)}{\partial
R}\right)
\\
&\quad-\dfrac{\im A(R)}{M} \dfrac{\partial}{\partial R}
\\
&\quad
-\dfrac{1}{2M}\dfrac{\partial^2}{\partial R^2},
\end{split}
\ee
where 
\be
\im\dfrac{\partial
A(R)}{\partial
R}=\left\langle\dfrac{\partial
\Phi_R}{\partial R}\middle\vert\dfrac{\partial\Phi_R}{\partial R}\right\rangle+
\left\langle \Phi_R\middle\vert
\dfrac{\partial^2 \Phi_R}{\partial R^2}\right\rangle,
\ee
Eq.~(\ref{eqapp:developed_nuclear_eq}) can be rewritten as follows,
\be\label{eqapp:first_appearance_cond_elec_ener}
\begin{split}
&\frac{\left(-\im\frac{\partial}{\partial
R}+A(R)\right)^2}{2M}\chi(R)
\\
&+\Bigg[\mathcal{E}^{\rm BO}(R)-\dfrac{A^2(R)}{2M}
+\dfrac{1}{2M}\left\langle\dfrac{\partial
\Phi_R}{\partial R}\middle\vert\dfrac{\partial\Phi_R}{\partial R}\right\rangle\Bigg]
\\
&\times\chi(R)=E\chi(R),
\end{split}
\ee
which, according to Eq.~(\ref{eq:expression_conditional_ener}), leads to the marginal nuclear Eq.~(\ref{eq:regular_marg_nuclear_eq}).\\

We now turn to the conditional electronic wavefunction $\Phi_R(r)$. The equation it fulfills is determined simply by dividing the original molecular Schrödinger Eq.~(\ref{eq:molecular_int_Schroedinger_eq}) by $\chi(R)$, which gives
\be
\begin{split}
&-\dfrac{1}{2M\chi(R)}\dfrac{\partial^2 \left[\chi(R)\Phi_R(r)\right]}{\partial R^2}
+\hat{H}^{\rm BO}(R)\Phi_R(r)
\\
&=E\Phi_R(r),
\end{split}
\ee
or, equivalently,
\be\label{eqapp:original_cond_eq_with_mol_E}
\begin{split}
&
\hat{H}^{\rm BO}(R)\Phi_R(r)
-\dfrac{1}{2M}\left(\dfrac{1}{\chi(R)}\dfrac{\partial^2 \chi(R)}{\partial R^2}\right)\Phi_R(r)
\\
&-\dfrac{1}{M}\dfrac{1}{\chi(R)}\dfrac{\partial\chi(R)}{\partial R}\dfrac{\partial\Phi_R(r)}{\partial R}
\\
&
-\dfrac{1}{2M}\dfrac{\partial^2 \Phi_R(r)}{\partial R^2}
=E\Phi_R(r).
\end{split}
\ee
In order to disentangle the above equation from that of the nuclei, we can use the following relation that we obtain by dividing the marginal nuclear Eq.~(\ref{eq:regular_marg_nuclear_eq}) by $\chi(R)$,
\be
E=\mathcal{E}(R)+\dfrac{1}{\chi(R)}\frac{\left(-\im\frac{\partial}{\partial
R}+A(R)\right)^2}{2M}\chi(R).
\ee
When inserted into Eq.~(\ref{eqapp:original_cond_eq_with_mol_E}) it gives
\be
\begin{split}
&
\hat{H}^{\rm BO}(R)\Phi_R(r)
-\dfrac{1}{2M}\left(\dfrac{1}{\chi(R)}\dfrac{\partial^2 \chi(R)}{\partial R^2}\right)\Phi_R(r)
\\
&-\dfrac{1}{M}\dfrac{1}{\chi(R)}\dfrac{\partial\chi(R)}{\partial R}\dfrac{\partial\Phi_R(r)}{\partial R}
\\
&
-\dfrac{1}{2M}\dfrac{\partial^2 \Phi_R(r)}{\partial R^2}
-\left[\dfrac{1}{\chi(R)}\frac{\left(-\im\frac{\partial}{\partial
R}+A(R)\right)^2}{2M}\chi(R)\right]\Phi_R(r)
\\
&=\mathcal{E}(R)\Phi_R(r),
\end{split}
\ee
or, equivalently [see the expansion in Eq.~(\ref{eqapp:expansion_momentum_plus_A_square})],
\be\label{eqapp:cond_elec_eq_developped_coupling_pot}
\begin{split}
&
\hat{H}^{\rm BO}(R)\Phi_R(r)
-\dfrac{1}{M}\dfrac{1}{\chi(R)}\dfrac{\partial\chi(R)}{\partial R}\dfrac{\partial\Phi_R(r)}{\partial R}
\\
&
-\dfrac{1}{2M}\dfrac{\partial^2 \Phi_R(r)}{\partial R^2}
\\
&-\Bigg[\dfrac{A^2(R)}{2M}-\dfrac{\im}{2M} \left(\dfrac{\partial
A(R)}{\partial
R}\right)
\\
&
-\dfrac{\im A(R)}{M\chi(R)} \dfrac{\partial \chi(R)}{\partial R}
\Bigg]\Phi_R(r)
=\mathcal{E}(R)\Phi_R(r).
\end{split}
\ee
By analogy with Eq.~(\ref{eqapp:expansion_momentum_plus_A_square}), we can rewrite the above equation in a more compact way as follows (note the change of sign $A(R)\rightarrow -A(R)$ in the first term on the right-hand side of the equation below), provided that two corrections (second and third terms on the right-hand side) and two missing terms (last contribution on the right-hand side) are added, which gives
\be
\begin{split}
&\left(\mathcal{E}(R)-\hat{H}^{\rm BO}(R)\right)\Phi_R(r)
\\
&=\frac{\left(-\im\frac{\partial}{\partial
R}-A(R)\right)^2}{2M}\Phi_R(r)-\dfrac{A^2(R)}{M}\Phi_R(r)
\\
&\quad -\dfrac{\im A(R)}{M}\dfrac{\partial\Phi_R(r)}{\partial R}
\\
&\quad-\dfrac{\im}{M}\dfrac{1}{\chi(R)}\dfrac{\partial\chi(R)}{\partial R}\left(-\im \dfrac{\partial}{\partial R}-A(R)\right)\Phi_R(r),
\end{split}
\ee
or, equivalently,
\be
\begin{split}
&\left[\hat{H}^{\rm BO}(R)
+\frac{\left(-\im\frac{\partial}{\partial
R}-A(R)\right)^2}{2M}\right]\Phi_R(r)
\\
&+\dfrac{A(R)}{M}\left(-\im \dfrac{\partial}{\partial R}-A(R)\right)\Phi_R(r)
\\
&-\dfrac{\im}{M}\dfrac{1}{\chi(R)}\dfrac{\partial\chi(R)}{\partial R}\left(-\im \dfrac{\partial}{\partial R}-A(R)\right)\Phi_R(r)
\\
&=\mathcal{E}(R)\Phi_R(r).
\end{split}
\ee
Thus, we recover the expected conditional electronic Eq.~(\ref{eq:XF_conditional_eq}), where the inverse of the marginal wave function is written as $\dfrac{1}{\chi(R)}=\dfrac{\chi(R)^*}{\vert \chi(R)\vert^2}$.

\section{Perturbative solutions to the ``first-order'' approximation}\label{app:PT}

For analysis purposes, we solve in this appendix the approximate conditional one-electron KS Eq.~(\ref{eq:one_elec_XF-based_KS_eq}) or, equivalently, the following non-interacting many-electron KS-like equation,  
\rev{\be\label{eqapp:eq_to_be_solved_in_PT}
\begin{split}
\left[\hat{T}_{\rm e}+\hat{V}^{\rm KS}_{\rm
ne}(R)+u^{\rm coup(1)}_{\rm ne,GI}
.\left(-\im\dfrac{\partial}{\partial
R}-A^{\rm KS}_{\mu}(R)\right)\right] & \Phi^\mu_R(r)
\\
=\underline{\mathcal{E}}^\mu(R) \; & \Phi^\mu_R(r),
\end{split}
\ee}
in perturbation theory.  
Note that Eq.~(\ref{eqapp:eq_to_be_solved_in_PT}) corresponds to Eq.~(\ref{eq:KS_cond_eq_no_second_order_deriv_scalar_form}) from which the electronically ``constant'' (\ie, $r$-independent) potential energy contribution $V^{\rm KS}_{\rm nn}(R)$ has been removed. Consequently, the total energy $\underline{\mathcal{E}}^{\mu}(R)$ in the right-hand side of Eq.~(\ref{eqapp:eq_to_be_solved_in_PT}) corresponds to the sum of the occupied-in-$\Phi^{\mu}_R$ orbital energies [first term on the right-hand side of Eq.~(\ref{eq:ener_state_mu_of_R})].\\

Let us now introduce the auxiliary equation   
\rev{\be\label{eqapp:aux_eq_in_alpha}
\begin{split}
&\left[\hat{T}_{\rm e}+\hat{V}^{\rm KS}_{\rm
ne}(R)+\alpha\,u^{\rm coup(1)}_{\rm ne,GI}
.\left(-\im\dfrac{\partial}{\partial
R}-A^{\rm KS}_{\mu,\alpha}(R)\right)\right]\Phi^{\mu,\alpha}_R(r)
\\
&=\underline{\mathcal{E}}^{\mu,\alpha}(R)\Phi^{\mu,\alpha}_R(r),
\end{split}
\ee}
where the coupling constant $\alpha$ varies in the range $0\leq \alpha\leq 1$. We consider the following Taylor expansions in $\alpha$,
\begin{subequations}
\begin{align}
\Phi^{\mu,\alpha}_R(r)&=\Phi^{\mu(0)}_R(r)+\alpha\Phi^{\mu(1)}_R(r)+\ldots,
\\
\underline{\mathcal{E}}^{\mu,\alpha}(R)&=\underline{\mathcal{E}}^{\mu(0)}(R)+\alpha\underline{\mathcal{E}}^{\mu(1)}(R)+\alpha^2\underline{\mathcal{E}}^{\mu(2)}(R)+\ldots,
\end{align}
\end{subequations}
where the first-order correction to the wavefunction can be expanded as follows,
\be\label{eqapp:1st_order_wf_exp}
\Phi^{\mu(1)}_R(r)=\sum_{\nu\neq\mu}C^{(1)}_{\mu\nu}(R)\Phi^{\nu(0)}_R(r).
\ee
\rev{and the vector potential reads}
\rev{\be
A^{\rm KS}_{\mu,\alpha}(R) = \left\langle \Phi^{\mu,\alpha}_R \middle| - \im \frac{\partial \Phi^{\mu,\alpha}_R}{\partial R}  \right\rangle_r 
\ee}
For any geometry $R$, the complete set $\{\Phi^{\mu(0)}_R\}$ of solutions to the unperturbed ($\alpha=0$) problem,
\be
\left(\hat{T}_{\rm e}+\hat{V}^{\rm KS}_{\rm
ne}(R)\right)\Phi^{\mu(0)}_R(r)=\underline{\mathcal{E}}^{\mu(0)}(R)\Phi^{\mu(0)}_R(r),
\ee
consists of {\it real-valued} and orthonormal (single-determinant) wavefunctions, \ie,
\be\label{eq:orthonormal_unpert_basis}
\langle \Phi^{\mu(0)}_R \vert \Phi^{\nu(0)}_R\rangle=\delta_{\mu\nu},\;\forall R, 
\ee
which implies
\be\label{eqapp:diag_NAC_zero}
\rev{A^{\rm KS}_{\mu,(0)}(R) =} \left\langle \Phi^{\mu(0)}_R\middle\vert \rev{ -\im } \dfrac{\partial \Phi^{\mu(0)}_R}{\partial R} \right\rangle=0,\;\forall\mu,
\ee
and
\be\label{eqapp:offdiag_NACs}
\left\langle \Phi^{\mu(0)}_R\middle\vert \dfrac{\partial \Phi^{\nu(0)}_R}{\partial R} \right\rangle
\overset{\mu\neq \nu}{=}
-\left\langle \Phi^{\nu(0)}_R\middle\vert \dfrac{\partial \Phi^{\mu(0)}_R}{\partial R} \right\rangle
.
\ee
The above equality translates the non-Hermiticity of the perturbation operator in this context [see Eq.~(\ref{eqapp:aux_eq_in_alpha})], unlike in regular perturbation theory. Once the perturbation expansions have been determined (see below), we approximate the solutions to Eq.~(\ref{eqapp:eq_to_be_solved_in_PT}) by applying the Taylor expansions up to $\alpha=1$, \ie,
\be
\Phi^{\mu}_R(r)\approx\Phi^{\mu(0)}_R(r)+\Phi^{\mu(1)}_R(r)+\ldots
\ee
and
\be
\underline{\mathcal{E}}^{\mu}(R)\approx \underline{\mathcal{E}}^{\mu(0)}(R)+\underline{\mathcal{E}}^{\mu(1)}(R)+\underline{\mathcal{E}}^{\mu(2)}(R)+\ldots
\ee
The first-order corrections are determined by keeping linear terms in $\alpha$ only (taking care to expand the vector potential in orders of $\alpha$ aswell), in Eq.~(\ref{eqapp:aux_eq_in_alpha}), thus leading to   
\be\label{eqapp:1st_order_eq}
\begin{split}
&\sum_{\nu\neq\mu}C^{(1)}_{\mu\nu}(R)\underline{\mathcal{E}}^{\nu(0)}(R)\Phi^{\nu(0)}_R(r)\rev{-\im} \,u^{\rm coup(1)}_{\rm
ne}(R)\dfrac{\partial \Phi^{\mu(0)}_R(r)}{\partial R}
\\
&=\underline{\mathcal{E}}^{\mu(1)}(R)\Phi^{\mu(0)}_R(r)+\underline{\mathcal{E}}^{\mu(0)}(R)\sum_{\nu\neq\mu}C^{(1)}_{\mu\nu}(R)\Phi^{\nu(0)}_R(r).
\end{split}
\ee

Projecting Eq.~(\ref{eqapp:1st_order_eq}) onto $\Phi^{\mu(0)}_R$ gives the first-order correction to the energy, which, according to Eq.~(\ref{eqapp:diag_NAC_zero}), turns out to be zero: 
\be\label{eqapp=zero_1st_order_corr_ener}
\underline{\mathcal{E}}^{\mu(1)}(R)= \rev{-\im} \, u^{\rm coup(1)}_{\rm
ne}(R)\left\langle \Phi^{\mu(0)}_R\middle\vert \dfrac{\partial \Phi^{\mu(0)}_R}{\partial R} \right\rangle=0.
\ee
Projecting Eq.~(\ref{eqapp:1st_order_eq}) onto $\Phi^{\nu(0)}_R$ ($\nu\neq \mu$) leads, on the other hand, to the first-order expansion of the wavefunction, {\it via} Eq.~(\ref{eqapp:1st_order_wf_exp}): 
\be\label{eqapp:1st_order_coeff_exp}
C^{(1)}_{\mu\nu}(R)=u^{\rm coup(1)}_{\rm
ne}(R)\dfrac{\left\langle \Phi^{\nu(0)}_R\middle\vert \rev{-\im} \, \dfrac{\partial \Phi^{\mu(0)}_R}{\partial R} \right\rangle}{\underline{\mathcal{E}}^{\mu(0)}(R)-\underline{\mathcal{E}}^{\nu(0)}(R)}.
\ee
Let us stress that the above coefficient matrix element is {\it symmetric}, unlike in conventional (Hermitian) perturbation theory, due to the anti-symmetry relation of Eq.~(\ref{eqapp:offdiag_NACs}):
\be
C^{(1)}_{\mu\nu}(R)\overset{\mu\neq \nu}{=}C^{(1)}_{\nu\mu}(R).
\ee
An important and expected implication is that the orthogonality between two different solutions is lost beyond the zeroth order. Indeed, at first order, we have
\be
\langle \Phi^{\kappa}_R  \vert \Phi^{\mu}_R \rangle \overset{\mu\neq \kappa}{\approx} C^{(1)}_{\mu\kappa}(R)+C^{(1)}_{\kappa\mu}(R)=2C^{(1)}_{\mu\kappa}(R),
\ee
or, more explicitly,
\be\label{eq:non_ortho_solutions}
\langle \Phi^{\kappa}_R  \vert \Phi^{\mu}_R \rangle \overset{\mu\neq \kappa}{\approx} 
2 u^{\rm coup(1)}_{\rm
ne}(R)\dfrac{\left\langle \Phi^{\kappa(0)}_R\middle\vert \rev{-\im} \, \dfrac{\partial \Phi^{\mu(0)}_R}{\partial R} \right\rangle}{\underline{\mathcal{E}}^{\mu(0)}(R)-\underline{\mathcal{E}}^{\kappa(0)}(R)}.
\ee
\rev{Until now no assumption about the gauge was made in the analysis. Note however that upon fixing the gauge by setting the KS nuclear wavefunction to be real, we have
\be
u^{\rm coup(1)}_{\rm
ne}(R) = - \frac{\im}{2 M} \frac{1}{\Gamma_0(R)} \frac{\partial \Gamma_0(R)}{\partial R} + A^{\rm KS}(R)
\ee
and if, additionally, the molecular wavefunction can be chosen real, the vector potential has to vanish everywhere
\be \label{eq:KS_1sto_u_nocurrent}
u^{\rm coup(1)}_{\rm
ne}(R) = - \frac{\im}{2 M} \frac{1}{\Gamma_0(R)} \frac{\partial \Gamma_0(R)}{\partial R}
\ee
and the first order coefficient matrix elements become real. This is the situation of our model system study}

\section{Computation of KS coefficients for the model system study} \label{sec:appHubKS}

Given a set of KS electronic coefficients, we determine their geometrical derivative and the KS potential through the ``first-order'' approximation of Eq.~(\ref{eq:KS_cond_eq_no_second_order_deriv_scalar_form}). \rev{First, we can freely choose the total molecular wavefunction to be real as it describes a bound state in a system with no electronic degeneracy. By setting the gauge so that the nuclear wavefunction is also real [see Eq.~\eqref{eq:gauge}], it follows that the electronic wavefunction must be real as well.} The normalization of the conditional wavefunction ensures $\langle \Phi^0_R| \nabla_R \Phi^0_R \rangle_r=0$, \rev{and thus the vector potential vanishes. We thus rewrite Eq.~(\ref{eq:KS_cond_eq_no_second_order_deriv_scalar_form}) as}
\begin{equation}
    \left[\hat{h}^{(0)}(R)- \mathcal{E}^0(R)\right]\Phi^0_R(r)
 = \rev{\im} \, u^{\rm coup(1)}_{\rm
ne}(R)\frac{\partial\Phi^0_R(r)}{\partial R}
\end{equation}
\rev{where $\mathcal{E}^0(R)$ is directly given by the expectation value of $\hat{h}^{(0)}(R)$ in our gauge} [see Eq.~\eqref{eq:KS_cond_eq_no_second_order_deriv_scalar_form}] \rev{and $u^{\rm coup(1)}_{\rm
ne}(R)$ is given by Eq.~\eqref{eq:KS_1sto_u_nocurrent}}. Dependence on the KS nuclear potential and $u_{\rm ne}^{\rm coup (0)}(R)$ thus simplifies out on the left-hand-side [see Eqs.~(\ref{eq:substitution_HBO_hmKS},\ref{eq:h0_def_mKS_plus_u0})]: 
\begin{equation}
\begin{split}
        \left[\hat{T}_{\rm e}+\hat{V}^{\rm KS}_{\rm ne}(R) - \langle \Phi^0_R | \hat{T}_{\rm e}+\hat{V}^{\rm KS}_{\rm ne}(R) | \Phi^0_R \rangle_r \right] \Phi^0_R(r) 
        \\
        = \rev{\im} \, u^{\rm coup(1)}_{\rm ne}(R)\frac{\partial\Phi^0_R(r)}{\partial R}
\end{split}
\end{equation}
Hence, the first-order approximation to the KS coefficients' geometrical derivatives can be computed at any $R$ from the coefficients themselves, the KS electronic potential, the nuclear density and its gradient (appearing in $u^{\rm coup(1)}_{\rm ne}(R)$). For the Hubbard dimer under study, the expression reads more explicitly
\begin{equation}
        \frac{\partial {\bf C}(R)}{\partial R} = \rev{\im} \frac{\Big[{\bm h}_s(R)-{\bm 1} ({\bf C}^\dagger.{\bm h}_s.{\bf C})(R) \Big].{\bf C}(R)}{u^{\rm coup(1)}_{\rm
ne}(R)}
\end{equation}
with ${\bf C}(R)=\big(C_1(R), C_2(R),C_3(R)\big)$ the vector of KS coefficients defining the conditional wavefunction $\displaystyle \Phi^0_R = \sum_{i=1}^3 C_i(R) \, \Phi_i$ and the KS Hamiltonian matrix
\begin{equation}
    {\bm h}_s(R) = \begin{bmatrix}
-\Delta v_s(R) & -\sqrt{2} \, t(R) & 0
\\
-\sqrt{2} \, t(R) & 0 & -\sqrt{2} \, t(R) 
\\
0 & -\sqrt{2} \, t(R) & \Delta v_s(R)
\end{bmatrix}
\end{equation}

From the above equations, it is clear that this procedure should not be applied in the vicinity of a vanishing nuclear density gradient, as $u^{\rm coup(1)}$ goes to 0. However, in such a case, Eq.~(\ref{eq:KS_cond_eq_no_second_order_deriv_scalar_form}) simplifies to a standard local-in-$R$ eigenvalue problem
\begin{equation}
     \hat{h}^{(0)}(R)\,\Phi^0_R(r)
 = \mathcal{E}^0(R)\,\Phi^0_R(r),
\end{equation}
so that no computation of geometrical derivative would be needed.

\clearpage

\bibliography{biblio}

\newcommand{\Aa}[0]{Aa}
\begin{thebibliography}{50}%
\makeatletter
\providecommand \@ifxundefined [1]{%
 \@ifx{#1\undefined}
}%
\providecommand \@ifnum [1]{%
 \ifnum #1\expandafter \@firstoftwo
 \else \expandafter \@secondoftwo
 \fi
}%
\providecommand \@ifx [1]{%
 \ifx #1\expandafter \@firstoftwo
 \else \expandafter \@secondoftwo
 \fi
}%
\providecommand \natexlab [1]{#1}%
\providecommand \enquote  [1]{``#1''}%
\providecommand \bibnamefont  [1]{#1}%
\providecommand \bibfnamefont [1]{#1}%
\providecommand \citenamefont [1]{#1}%
\providecommand \href@noop [0]{\@secondoftwo}%
\providecommand \href [0]{\begingroup \@sanitize@url \@href}%
\providecommand \@href[1]{\@@startlink{#1}\@@href}%
\providecommand \@@href[1]{\endgroup#1\@@endlink}%
\providecommand \@sanitize@url [0]{\catcode `\\12\catcode `\$12\catcode
  `\&12\catcode `\#12\catcode `\^12\catcode `\_12\catcode `\%12\relax}%
\providecommand \@@startlink[1]{}%
\providecommand \@@endlink[0]{}%
\providecommand \url  [0]{\begingroup\@sanitize@url \@url }%
\providecommand \@url [1]{\endgroup\@href {#1}{\urlprefix }}%
\providecommand \urlprefix  [0]{URL }%
\providecommand \Eprint [0]{\href }%
\providecommand \doibase [0]{http://dx.doi.org/}%
\providecommand \selectlanguage [0]{\@gobble}%
\providecommand \bibinfo  [0]{\@secondoftwo}%
\providecommand \bibfield  [0]{\@secondoftwo}%
\providecommand \translation [1]{[#1]}%
\providecommand \BibitemOpen [0]{}%
\providecommand \bibitemStop [0]{}%
\providecommand \bibitemNoStop [0]{.\EOS\space}%
\providecommand \EOS [0]{\spacefactor3000\relax}%
\providecommand \BibitemShut  [1]{\csname bibitem#1\endcsname}%
\let\auto@bib@innerbib\@empty
\bibitem [{\citenamefont {Hohenberg}\ and\ \citenamefont
  {Kohn}(1964)}]{Hohenberg1964}%
  \BibitemOpen
  \bibfield  {author} {\bibinfo {author} {\bibfnamefont {P.}~\bibnamefont
  {Hohenberg}}\ and\ \bibinfo {author} {\bibfnamefont {W.}~\bibnamefont
  {Kohn}},\ }\href {\doibase 10.1103/PhysRev.136.B864} {\bibfield  {journal}
  {\bibinfo  {journal} {Phys. Rev.}\ }\textbf {\bibinfo {volume} {136}},\
  \bibinfo {pages} {B864} (\bibinfo {year} {1964})}\BibitemShut {NoStop}%
\bibitem [{\citenamefont {Kohn}\ and\ \citenamefont {Sham}(1965)}]{Kohn1965}%
  \BibitemOpen
  \bibfield  {author} {\bibinfo {author} {\bibfnamefont {W.}~\bibnamefont
  {Kohn}}\ and\ \bibinfo {author} {\bibfnamefont {L.~J.}\ \bibnamefont
  {Sham}},\ }\href {\doibase 10.1103/PhysRev.140.A1133} {\bibfield  {journal}
  {\bibinfo  {journal} {Phys. Rev.}\ }\textbf {\bibinfo {volume} {140}},\
  \bibinfo {pages} {A1133} (\bibinfo {year} {1965})}\BibitemShut {NoStop}%
\bibitem [{\citenamefont {Burke}(2012)}]{Burke2012perspectiveDFA}%
  \BibitemOpen
  \bibfield  {author} {\bibinfo {author} {\bibfnamefont {K.}~\bibnamefont
  {Burke}},\ }\href {\doibase 10.1063/1.4704546} {\bibfield  {journal}
  {\bibinfo  {journal} {The Journal of Chemical Physics}\ }\textbf {\bibinfo
  {volume} {136}},\ \bibinfo {pages} {150901} (\bibinfo {year} {2012})},\
  \Eprint
  {http://arxiv.org/abs/https://pubs.aip.org/aip/jcp/article-pdf/doi/10.1063/1.4704546/19858608/150901\_1\_1.4704546.pdf}
  {https://pubs.aip.org/aip/jcp/article-pdf/doi/10.1063/1.4704546/19858608/150901\_1\_1.4704546.pdf}
  \BibitemShut {NoStop}%
\bibitem [{\citenamefont {Teale}\ \emph {et~al.}(2022)\citenamefont {Teale},
  \citenamefont {Helgaker}, \citenamefont {Savin}, \citenamefont {Adamo},
  \citenamefont {Aradi}, \citenamefont {Arbuznikov}, \citenamefont {Ayers},
  \citenamefont {Baerends}, \citenamefont {Barone}, \citenamefont
  {Calaminici},\ and\ \citenamefont {{et {\it al.}}}}]{Teale2022_DFT_exchange}%
  \BibitemOpen
  \bibfield  {author} {\bibinfo {author} {\bibfnamefont {A.~M.}\ \bibnamefont
  {Teale}}, \bibinfo {author} {\bibfnamefont {T.}~\bibnamefont {Helgaker}},
  \bibinfo {author} {\bibfnamefont {A.}~\bibnamefont {Savin}}, \bibinfo
  {author} {\bibfnamefont {C.}~\bibnamefont {Adamo}}, \bibinfo {author}
  {\bibfnamefont {B.}~\bibnamefont {Aradi}}, \bibinfo {author} {\bibfnamefont
  {A.~V.}\ \bibnamefont {Arbuznikov}}, \bibinfo {author} {\bibfnamefont
  {P.~W.}\ \bibnamefont {Ayers}}, \bibinfo {author} {\bibfnamefont {E.~J.}\
  \bibnamefont {Baerends}}, \bibinfo {author} {\bibfnamefont {V.}~\bibnamefont
  {Barone}}, \bibinfo {author} {\bibfnamefont {P.}~\bibnamefont {Calaminici}},
  \ and\ \bibinfo {author} {\bibnamefont {{et {\it al.}}}},\ }\href {\doibase
  10.1039/D2CP02827A} {\bibfield  {journal} {\bibinfo  {journal} {Phys. Chem.
  Chem. Phys.}\ }\textbf {\bibinfo {volume} {24}},\ \bibinfo {pages} {28700}
  (\bibinfo {year} {2022})}\BibitemShut {NoStop}%
\bibitem [{\citenamefont {Runge}\ and\ \citenamefont
  {Gross}(1984)}]{runge1984density}%
  \BibitemOpen
  \bibfield  {author} {\bibinfo {author} {\bibfnamefont {E.}~\bibnamefont
  {Runge}}\ and\ \bibinfo {author} {\bibfnamefont {E.~K.}\ \bibnamefont
  {Gross}},\ }\href {https://doi.org/10.1103/PhysRevLett.52.997} {\bibfield
  {journal} {\bibinfo  {journal} {Phys. Rev. Lett.}\ }\textbf {\bibinfo
  {volume} {52}},\ \bibinfo {pages} {997} (\bibinfo {year} {1984})}\BibitemShut
  {NoStop}%
\bibitem [{\citenamefont {Casida}\ and\ \citenamefont
  {Huix-Rotllant}(2012)}]{Casida_tddft_review_2012}%
  \BibitemOpen
  \bibfield  {author} {\bibinfo {author} {\bibfnamefont {M.}~\bibnamefont
  {Casida}}\ and\ \bibinfo {author} {\bibfnamefont {M.}~\bibnamefont
  {Huix-Rotllant}},\ }\href
  {https://doi.org/10.1146/annurev-physchem-032511-143803} {\bibfield
  {journal} {\bibinfo  {journal} {Annu. Rev. Phys. Chem.}\ }\textbf {\bibinfo
  {volume} {63}},\ \bibinfo {pages} {287} (\bibinfo {year} {2012})}\BibitemShut
  {NoStop}%
\bibitem [{\citenamefont {Send}\ and\ \citenamefont
  {Furche}(2010)}]{Send10_First-order}%
  \BibitemOpen
  \bibfield  {author} {\bibinfo {author} {\bibfnamefont {R.}~\bibnamefont
  {Send}}\ and\ \bibinfo {author} {\bibfnamefont {F.}~\bibnamefont {Furche}},\
  }\href {\doibase 10.1063/1.3292571} {\bibfield  {journal} {\bibinfo
  {journal} {The Journal of Chemical Physics}\ }\textbf {\bibinfo {volume}
  {132}},\ \bibinfo {pages} {044107} (\bibinfo {year} {2010})}\BibitemShut
  {NoStop}%
\bibitem [{\citenamefont {Ou}\ \emph {et~al.}(2015)\citenamefont {Ou},
  \citenamefont {Bellchambers}, \citenamefont {Furche},\ and\ \citenamefont
  {Subotnik}}]{Ou15_First-order}%
  \BibitemOpen
  \bibfield  {author} {\bibinfo {author} {\bibfnamefont {Q.}~\bibnamefont
  {Ou}}, \bibinfo {author} {\bibfnamefont {G.~D.}\ \bibnamefont
  {Bellchambers}}, \bibinfo {author} {\bibfnamefont {F.}~\bibnamefont
  {Furche}}, \ and\ \bibinfo {author} {\bibfnamefont {J.~E.}\ \bibnamefont
  {Subotnik}},\ }\href {\doibase 10.1063/1.4906941} {\bibfield  {journal}
  {\bibinfo  {journal} {The Journal of Chemical Physics}\ }\textbf {\bibinfo
  {volume} {142}},\ \bibinfo {pages} {064114} (\bibinfo {year}
  {2015})}\BibitemShut {NoStop}%
\bibitem [{\citenamefont {Wang}\ \emph {et~al.}(2021)\citenamefont {Wang},
  \citenamefont {Wu},\ and\ \citenamefont {Liu}}]{Wang21_NAC-TDDFT}%
  \BibitemOpen
  \bibfield  {author} {\bibinfo {author} {\bibfnamefont {Z.}~\bibnamefont
  {Wang}}, \bibinfo {author} {\bibfnamefont {C.}~\bibnamefont {Wu}}, \ and\
  \bibinfo {author} {\bibfnamefont {W.}~\bibnamefont {Liu}},\ }\href {\doibase
  10.1021/acs.accounts.1c00312} {\bibfield  {journal} {\bibinfo  {journal}
  {Accounts of Chemical Research}\ }\textbf {\bibinfo {volume} {54}},\ \bibinfo
  {pages} {3288} (\bibinfo {year} {2021})}\BibitemShut {NoStop}%
\bibitem [{\citenamefont {Domcke}\ \emph {et~al.}(2004)\citenamefont {Domcke},
  \citenamefont {Yarkony},\ and\ \citenamefont {K\"oppel}}]{dom04}%
  \BibitemOpen
  \bibinfo {editor} {\bibfnamefont {W.}~\bibnamefont {Domcke}}, \bibinfo
  {editor} {\bibfnamefont {D.~R.}\ \bibnamefont {Yarkony}}, \ and\ \bibinfo
  {editor} {\bibfnamefont {H.}~\bibnamefont {K\"oppel}},\ eds.,\ \enquote
  {\bibinfo {title} {Conical intersections: Electronic structure, dynamics \&
  spectroscopy},}\ \ (\bibinfo  {publisher} {World Scientific},\ \bibinfo
  {address} {Singapore},\ \bibinfo {year} {2004})\BibitemShut {NoStop}%
\bibitem [{\citenamefont {Baer}(2006)}]{bae06}%
  \BibitemOpen
  \bibfield  {author} {\bibinfo {author} {\bibfnamefont {M.}~\bibnamefont
  {Baer}},\ }\enquote {\bibinfo {title} {Beyond born-oppenheimer: Electronic
  nonadiabatic coupling terms and conical intersections},}\ \ (\bibinfo
  {publisher} {Wiley},\ \bibinfo {address} {Hoboken, {NJ}},\ \bibinfo {year}
  {2006})\BibitemShut {NoStop}%
\bibitem [{\citenamefont {Domcke}\ \emph {et~al.}(2011)\citenamefont {Domcke},
  \citenamefont {Yarkony},\ and\ \citenamefont {K\"oppel}}]{dom11}%
  \BibitemOpen
  \bibinfo {editor} {\bibfnamefont {W.}~\bibnamefont {Domcke}}, \bibinfo
  {editor} {\bibfnamefont {D.~R.}\ \bibnamefont {Yarkony}}, \ and\ \bibinfo
  {editor} {\bibfnamefont {H.}~\bibnamefont {K\"oppel}},\ eds.,\ \enquote
  {\bibinfo {title} {Conical intersections: Theory, computation and
  experiment},}\ \ (\bibinfo  {publisher} {World Scientific},\ \bibinfo
  {address} {Singapore},\ \bibinfo {year} {2011})\BibitemShut {NoStop}%
\bibitem [{\citenamefont {Lasorne}\ \emph {et~al.}(2011)\citenamefont
  {Lasorne}, \citenamefont {Worth},\ and\ \citenamefont {Robb}}]{las11:460}%
  \BibitemOpen
  \bibfield  {author} {\bibinfo {author} {\bibfnamefont {B.}~\bibnamefont
  {Lasorne}}, \bibinfo {author} {\bibfnamefont {G.~A.}\ \bibnamefont {Worth}},
  \ and\ \bibinfo {author} {\bibfnamefont {M.~A.}\ \bibnamefont {Robb}},\
  }\href {\doibase https://doi.org/10.1002/wcms.26} {\bibfield  {journal}
  {\bibinfo  {journal} {WIREs Computational Molecular Science}\ }\textbf
  {\bibinfo {volume} {1}},\ \bibinfo {pages} {460} (\bibinfo {year}
  {2011})}\BibitemShut {NoStop}%
\bibitem [{\citenamefont {Fromager}\ and\ \citenamefont
  {Lasorne}(2024)}]{Fromager_2024_Density}%
  \BibitemOpen
  \bibfield  {author} {\bibinfo {author} {\bibfnamefont {E.}~\bibnamefont
  {Fromager}}\ and\ \bibinfo {author} {\bibfnamefont {B.}~\bibnamefont
  {Lasorne}},\ }\href {\doibase 10.1088/2516-1075/ad45d5} {\bibfield  {journal}
  {\bibinfo  {journal} {Electronic Structure}\ }\textbf {\bibinfo {volume}
  {6}},\ \bibinfo {pages} {025002} (\bibinfo {year} {2024})}\BibitemShut
  {NoStop}%
\bibitem [{\citenamefont {Kreibich}\ and\ \citenamefont
  {Gross}(2001)}]{Kreibich2001}%
  \BibitemOpen
  \bibfield  {author} {\bibinfo {author} {\bibfnamefont {T.}~\bibnamefont
  {Kreibich}}\ and\ \bibinfo {author} {\bibfnamefont {E.~K.~U.}\ \bibnamefont
  {Gross}},\ }\href {\doibase 10.1103/PhysRevLett.86.2984} {\bibfield
  {journal} {\bibinfo  {journal} {Phys. Rev. Lett.}\ }\textbf {\bibinfo
  {volume} {86}},\ \bibinfo {pages} {2984} (\bibinfo {year}
  {2001})}\BibitemShut {NoStop}%
\bibitem [{\citenamefont {Gidopoulos}(1998)}]{Gidopoulos98}%
  \BibitemOpen
  \bibfield  {author} {\bibinfo {author} {\bibfnamefont {N.}~\bibnamefont
  {Gidopoulos}},\ }\href {\doibase 10.1103/PhysRevB.57.2146} {\bibfield
  {journal} {\bibinfo  {journal} {Phys. Rev. B}\ }\textbf {\bibinfo {volume}
  {57}},\ \bibinfo {pages} {2146} (\bibinfo {year} {1998})}\BibitemShut
  {NoStop}%
\bibitem [{\citenamefont {Butriy}\ \emph {et~al.}(2007)\citenamefont {Butriy},
  \citenamefont {Ebadi}, \citenamefont {de~Boeij}, \citenamefont {van
  Leeuwen},\ and\ \citenamefont {Gross}}]{Butriy07}%
  \BibitemOpen
  \bibfield  {author} {\bibinfo {author} {\bibfnamefont {O.}~\bibnamefont
  {Butriy}}, \bibinfo {author} {\bibfnamefont {H.}~\bibnamefont {Ebadi}},
  \bibinfo {author} {\bibfnamefont {P.~L.}\ \bibnamefont {de~Boeij}}, \bibinfo
  {author} {\bibfnamefont {R.}~\bibnamefont {van Leeuwen}}, \ and\ \bibinfo
  {author} {\bibfnamefont {E.~K.~U.}\ \bibnamefont {Gross}},\ }\href {\doibase
  10.1103/PhysRevA.76.052514} {\bibfield  {journal} {\bibinfo  {journal} {Phys.
  Rev. A}\ }\textbf {\bibinfo {volume} {76}},\ \bibinfo {pages} {052514}
  (\bibinfo {year} {2007})}\BibitemShut {NoStop}%
\bibitem [{\citenamefont {Kreibich}\ \emph {et~al.}(2008)\citenamefont
  {Kreibich}, \citenamefont {van Leeuwen},\ and\ \citenamefont
  {Gross}}]{Kreibich08}%
  \BibitemOpen
  \bibfield  {author} {\bibinfo {author} {\bibfnamefont {T.}~\bibnamefont
  {Kreibich}}, \bibinfo {author} {\bibfnamefont {R.}~\bibnamefont {van
  Leeuwen}}, \ and\ \bibinfo {author} {\bibfnamefont {E.~K.~U.}\ \bibnamefont
  {Gross}},\ }\href {\doibase 10.1103/PhysRevA.78.022501} {\bibfield  {journal}
  {\bibinfo  {journal} {Phys. Rev. A}\ }\textbf {\bibinfo {volume} {78}},\
  \bibinfo {pages} {022501} (\bibinfo {year} {2008})}\BibitemShut {NoStop}%
\bibitem [{\citenamefont {Chakraborty}\ \emph {et~al.}(2008)\citenamefont
  {Chakraborty}, \citenamefont {Pak},\ and\ \citenamefont
  {Hammes-Schiffer}}]{Chakraborty2008_Development}%
  \BibitemOpen
  \bibfield  {author} {\bibinfo {author} {\bibfnamefont {A.}~\bibnamefont
  {Chakraborty}}, \bibinfo {author} {\bibfnamefont {M.~V.}\ \bibnamefont
  {Pak}}, \ and\ \bibinfo {author} {\bibfnamefont {S.}~\bibnamefont
  {Hammes-Schiffer}},\ }\href {\doibase 10.1103/PhysRevLett.101.153001}
  {\bibfield  {journal} {\bibinfo  {journal} {Phys. Rev. Lett.}\ }\textbf
  {\bibinfo {volume} {101}},\ \bibinfo {pages} {153001} (\bibinfo {year}
  {2008})}\BibitemShut {NoStop}%
\bibitem [{\citenamefont {Mejía-Rodríguez}\ and\ \citenamefont {de~la
  Lande}(2019)}]{delaLande2019_Multicomponent}%
  \BibitemOpen
  \bibfield  {author} {\bibinfo {author} {\bibfnamefont {D.}~\bibnamefont
  {Mejía-Rodríguez}}\ and\ \bibinfo {author} {\bibfnamefont {A.}~\bibnamefont
  {de~la Lande}},\ }\href {\doibase 10.1063/1.5078596} {\bibfield  {journal}
  {\bibinfo  {journal} {The Journal of Chemical Physics}\ }\textbf {\bibinfo
  {volume} {150}},\ \bibinfo {pages} {174115} (\bibinfo {year}
  {2019})}\BibitemShut {NoStop}%
\bibitem [{\citenamefont {Xu}\ \emph {et~al.}(2023)\citenamefont {Xu},
  \citenamefont {Zhou}, \citenamefont {Blum}, \citenamefont {Li}, \citenamefont
  {Hammes-Schiffer},\ and\ \citenamefont {Kanai}}]{Xu2023_First-Principles}%
  \BibitemOpen
  \bibfield  {author} {\bibinfo {author} {\bibfnamefont {J.}~\bibnamefont
  {Xu}}, \bibinfo {author} {\bibfnamefont {R.}~\bibnamefont {Zhou}}, \bibinfo
  {author} {\bibfnamefont {V.}~\bibnamefont {Blum}}, \bibinfo {author}
  {\bibfnamefont {T.~E.}\ \bibnamefont {Li}}, \bibinfo {author} {\bibfnamefont
  {S.}~\bibnamefont {Hammes-Schiffer}}, \ and\ \bibinfo {author} {\bibfnamefont
  {Y.}~\bibnamefont {Kanai}},\ }\href {\doibase 10.1103/PhysRevLett.131.238002}
  {\bibfield  {journal} {\bibinfo  {journal} {Phys. Rev. Lett.}\ }\textbf
  {\bibinfo {volume} {131}},\ \bibinfo {pages} {238002} (\bibinfo {year}
  {2023})}\BibitemShut {NoStop}%
\bibitem [{\citenamefont {Requist}\ and\ \citenamefont
  {Gross}(2016)}]{Requist16_Exact}%
  \BibitemOpen
  \bibfield  {author} {\bibinfo {author} {\bibfnamefont {R.}~\bibnamefont
  {Requist}}\ and\ \bibinfo {author} {\bibfnamefont {E.~K.~U.}\ \bibnamefont
  {Gross}},\ }\href {\doibase 10.1103/PhysRevLett.117.193001} {\bibfield
  {journal} {\bibinfo  {journal} {Phys. Rev. Lett.}\ }\textbf {\bibinfo
  {volume} {117}},\ \bibinfo {pages} {193001} (\bibinfo {year}
  {2016})}\BibitemShut {NoStop}%
\bibitem [{\citenamefont {Li}\ \emph {et~al.}(2018)\citenamefont {Li},
  \citenamefont {Requist},\ and\ \citenamefont {Gross}}]{Li18_Density}%
  \BibitemOpen
  \bibfield  {author} {\bibinfo {author} {\bibfnamefont {C.}~\bibnamefont
  {Li}}, \bibinfo {author} {\bibfnamefont {R.}~\bibnamefont {Requist}}, \ and\
  \bibinfo {author} {\bibfnamefont {E.~K.~U.}\ \bibnamefont {Gross}},\ }\href
  {\doibase 10.1063/1.5011663} {\bibfield  {journal} {\bibinfo  {journal} {The
  Journal of Chemical Physics}\ }\textbf {\bibinfo {volume} {148}},\ \bibinfo
  {pages} {084110} (\bibinfo {year} {2018})}\BibitemShut {NoStop}%
\bibitem [{\citenamefont {Kapral}\ and\ \citenamefont
  {Ciccotti}(1999)}]{Kapral_Ciccotti99}%
  \BibitemOpen
  \bibfield  {author} {\bibinfo {author} {\bibfnamefont {R.}~\bibnamefont
  {Kapral}}\ and\ \bibinfo {author} {\bibfnamefont {G.}~\bibnamefont
  {Ciccotti}},\ }\href {\doibase 10.1063/1.478811} {\bibfield  {journal}
  {\bibinfo  {journal} {The Journal of Chemical Physics}\ }\textbf {\bibinfo
  {volume} {110}},\ \bibinfo {pages} {8919} (\bibinfo {year}
  {1999})}\BibitemShut {NoStop}%
\bibitem [{\citenamefont {Hunter}(1975)}]{Hunter1975_Conditional}%
  \BibitemOpen
  \bibfield  {author} {\bibinfo {author} {\bibfnamefont {G.}~\bibnamefont
  {Hunter}},\ }\href {\doibase https://doi.org/10.1002/qua.560090205}
  {\bibfield  {journal} {\bibinfo  {journal} {International Journal of Quantum
  Chemistry}\ }\textbf {\bibinfo {volume} {9}},\ \bibinfo {pages} {237}
  (\bibinfo {year} {1975})}\BibitemShut {NoStop}%
\bibitem [{\citenamefont {Abedi}\ \emph {et~al.}(2010)\citenamefont {Abedi},
  \citenamefont {Maitra},\ and\ \citenamefont {Gross}}]{Abedi10_EF}%
  \BibitemOpen
  \bibfield  {author} {\bibinfo {author} {\bibfnamefont {A.}~\bibnamefont
  {Abedi}}, \bibinfo {author} {\bibfnamefont {N.~T.}\ \bibnamefont {Maitra}}, \
  and\ \bibinfo {author} {\bibfnamefont {E.~K.~U.}\ \bibnamefont {Gross}},\
  }\href {\doibase 10.1103/PhysRevLett.105.123002} {\bibfield  {journal}
  {\bibinfo  {journal} {Phys. Rev. Lett.}\ }\textbf {\bibinfo {volume} {105}},\
  \bibinfo {pages} {123002} (\bibinfo {year} {2010})}\BibitemShut {NoStop}%
\bibitem [{\citenamefont {Abedi}\ \emph {et~al.}(2012)\citenamefont {Abedi},
  \citenamefont {Maitra},\ and\ \citenamefont {Gross}}]{Abedi12_Correlated}%
  \BibitemOpen
  \bibfield  {author} {\bibinfo {author} {\bibfnamefont {A.}~\bibnamefont
  {Abedi}}, \bibinfo {author} {\bibfnamefont {N.~T.}\ \bibnamefont {Maitra}}, \
  and\ \bibinfo {author} {\bibfnamefont {E.~K.~U.}\ \bibnamefont {Gross}},\
  }\href {\doibase 10.1063/1.4745836} {\bibfield  {journal} {\bibinfo
  {journal} {The Journal of Chemical Physics}\ }\textbf {\bibinfo {volume}
  {137}},\ \bibinfo {pages} {22A530} (\bibinfo {year} {2012})}\BibitemShut
  {NoStop}%
\bibitem [{\citenamefont {Gidopoulos}\ and\ \citenamefont
  {Gross}(2014)}]{Gidopoulos2014_Electronic}%
  \BibitemOpen
  \bibfield  {author} {\bibinfo {author} {\bibfnamefont {N.~I.}\ \bibnamefont
  {Gidopoulos}}\ and\ \bibinfo {author} {\bibfnamefont {E.~K.~U.}\ \bibnamefont
  {Gross}},\ }\href {\doibase 10.1098/rsta.2013.0059} {\bibfield  {journal}
  {\bibinfo  {journal} {Philosophical Transactions of the Royal Society A:
  Mathematical, Physical and Engineering Sciences}\ }\textbf {\bibinfo {volume}
  {372}},\ \bibinfo {pages} {20130059} (\bibinfo {year} {2014})}\BibitemShut
  {NoStop}%
\bibitem [{\citenamefont {Min}\ \emph {et~al.}(2015)\citenamefont {Min},
  \citenamefont {Agostini},\ and\ \citenamefont
  {Gross}}]{Min15_Coupled-Trajectory}%
  \BibitemOpen
  \bibfield  {author} {\bibinfo {author} {\bibfnamefont {S.~K.}\ \bibnamefont
  {Min}}, \bibinfo {author} {\bibfnamefont {F.}~\bibnamefont {Agostini}}, \
  and\ \bibinfo {author} {\bibfnamefont {E.~K.~U.}\ \bibnamefont {Gross}},\
  }\href {\doibase 10.1103/PhysRevLett.115.073001} {\bibfield  {journal}
  {\bibinfo  {journal} {Phys. Rev. Lett.}\ }\textbf {\bibinfo {volume} {115}},\
  \bibinfo {pages} {073001} (\bibinfo {year} {2015})}\BibitemShut {NoStop}%
\bibitem [{\citenamefont {Filatov}\ \emph {et~al.}(2018)\citenamefont
  {Filatov}, \citenamefont {Min},\ and\ \citenamefont
  {Kim}}]{Filatov18_Direct}%
  \BibitemOpen
  \bibfield  {author} {\bibinfo {author} {\bibfnamefont {M.}~\bibnamefont
  {Filatov}}, \bibinfo {author} {\bibfnamefont {S.~K.}\ \bibnamefont {Min}}, \
  and\ \bibinfo {author} {\bibfnamefont {K.~S.}\ \bibnamefont {Kim}},\ }\href
  {\doibase 10.1021/acs.jctc.8b00217} {\bibfield  {journal} {\bibinfo
  {journal} {Journal of Chemical Theory and Computation}\ }\textbf {\bibinfo
  {volume} {14}},\ \bibinfo {pages} {4499} (\bibinfo {year}
  {2018})}\BibitemShut {NoStop}%
\bibitem [{\citenamefont {Agostini}\ \emph {et~al.}(2016)\citenamefont
  {Agostini}, \citenamefont {Min}, \citenamefont {Abedi},\ and\ \citenamefont
  {Gross}}]{AMAG16}%
  \BibitemOpen
  \bibfield  {author} {\bibinfo {author} {\bibfnamefont {F.}~\bibnamefont
  {Agostini}}, \bibinfo {author} {\bibfnamefont {S.~K.}\ \bibnamefont {Min}},
  \bibinfo {author} {\bibfnamefont {A.}~\bibnamefont {Abedi}}, \ and\ \bibinfo
  {author} {\bibfnamefont {E.~K.~U.}\ \bibnamefont {Gross}},\ }\href {\doibase
  10.1021/acs.jctc.5b01180} {\bibfield  {journal} {\bibinfo  {journal} {J.
  Chem. Theory Comput.}\ }\textbf {\bibinfo {volume} {12}},\ \bibinfo {pages}
  {2127} (\bibinfo {year} {2016})}\BibitemShut {NoStop}%
\bibitem [{\citenamefont {Ha}\ \emph {et~al.}(2018)\citenamefont {Ha},
  \citenamefont {Lee},\ and\ \citenamefont {Min}}]{HLM18}%
  \BibitemOpen
  \bibfield  {author} {\bibinfo {author} {\bibfnamefont {J.-K.}\ \bibnamefont
  {Ha}}, \bibinfo {author} {\bibfnamefont {I.~S.}\ \bibnamefont {Lee}}, \ and\
  \bibinfo {author} {\bibfnamefont {S.~K.}\ \bibnamefont {Min}},\ }\href
  {\doibase 10.1021/acs.jpclett.8b00060} {\bibfield  {journal} {\bibinfo
  {journal} {J. Phys. Chem. Lett.}\ }\textbf {\bibinfo {volume} {9}},\ \bibinfo
  {pages} {1097} (\bibinfo {year} {2018})}\BibitemShut {NoStop}%
\bibitem [{\citenamefont {Ha}\ and\ \citenamefont {Min}(2022)}]{HM22}%
  \BibitemOpen
  \bibfield  {author} {\bibinfo {author} {\bibfnamefont {J.-K.}\ \bibnamefont
  {Ha}}\ and\ \bibinfo {author} {\bibfnamefont {S.~K.}\ \bibnamefont {Min}},\
  }\href {\doibase 10.1063/5.0084493} {\bibfield  {journal} {\bibinfo
  {journal} {J. Chem. Phys.}\ }\textbf {\bibinfo {volume} {156}},\ \bibinfo
  {pages} {174109} (\bibinfo {year} {2022})}\BibitemShut {NoStop}%
\bibitem [{\citenamefont {Vindel-Zandbergen}\ \emph {et~al.}(2022)\citenamefont
  {Vindel-Zandbergen}, \citenamefont {Matsika},\ and\ \citenamefont
  {Maitra}}]{VMM22}%
  \BibitemOpen
  \bibfield  {author} {\bibinfo {author} {\bibfnamefont {P.}~\bibnamefont
  {Vindel-Zandbergen}}, \bibinfo {author} {\bibfnamefont {S.}~\bibnamefont
  {Matsika}}, \ and\ \bibinfo {author} {\bibfnamefont {N.~T.}\ \bibnamefont
  {Maitra}},\ }\href {\doibase 10.1021/acs.jpclett.1c04132} {\bibfield
  {journal} {\bibinfo  {journal} {J. Phys. Chem. Lett.}\ }\textbf {\bibinfo
  {volume} {13}},\ \bibinfo {pages} {1785} (\bibinfo {year}
  {2022})}\BibitemShut {NoStop}%
\bibitem [{\citenamefont {Villaseco~Arribas}\ \emph {et~al.}(2022)\citenamefont
  {Villaseco~Arribas}, \citenamefont {Agostini},\ and\ \citenamefont
  {Maitra}}]{Villaseco22_Exact}%
  \BibitemOpen
  \bibfield  {author} {\bibinfo {author} {\bibfnamefont {E.}~\bibnamefont
  {Villaseco~Arribas}}, \bibinfo {author} {\bibfnamefont {F.}~\bibnamefont
  {Agostini}}, \ and\ \bibinfo {author} {\bibfnamefont {N.~T.}\ \bibnamefont
  {Maitra}},\ }\href {https://www.mdpi.com/1420-3049/27/13/4002} {\bibfield
  {journal} {\bibinfo  {journal} {Molecules}\ }\textbf {\bibinfo {volume} {27}}
  (\bibinfo {year} {2022})}\BibitemShut {NoStop}%
\bibitem [{\citenamefont {Dupuy}\ \emph {et~al.}(2024)\citenamefont {Dupuy},
  \citenamefont {Rikus},\ and\ \citenamefont {Maitra}}]{DRM24}%
  \BibitemOpen
  \bibfield  {author} {\bibinfo {author} {\bibfnamefont {L.}~\bibnamefont
  {Dupuy}}, \bibinfo {author} {\bibfnamefont {A.}~\bibnamefont {Rikus}}, \ and\
  \bibinfo {author} {\bibfnamefont {N.~T.}\ \bibnamefont {Maitra}},\ }\href
  {\doibase 10.1021/acs.jpclett.4c00115} {\bibfield  {journal} {\bibinfo
  {journal} {The Journal of Physical Chemistry Letters}\ }\textbf {\bibinfo
  {volume} {15}},\ \bibinfo {pages} {2643} (\bibinfo {year}
  {2024})}\BibitemShut {NoStop}%
\bibitem [{\citenamefont {Scherrer}\ \emph {et~al.}(2015)\citenamefont
  {Scherrer}, \citenamefont {Agostini}, \citenamefont {Sebastiani},
  \citenamefont {Gross},\ and\ \citenamefont {Vuilleumier}}]{NVPT_2015}%
  \BibitemOpen
  \bibfield  {author} {\bibinfo {author} {\bibfnamefont {A.}~\bibnamefont
  {Scherrer}}, \bibinfo {author} {\bibfnamefont {F.}~\bibnamefont {Agostini}},
  \bibinfo {author} {\bibfnamefont {D.}~\bibnamefont {Sebastiani}}, \bibinfo
  {author} {\bibfnamefont {E.~K.~U.}\ \bibnamefont {Gross}}, \ and\ \bibinfo
  {author} {\bibfnamefont {R.}~\bibnamefont {Vuilleumier}},\ }\href {\doibase
  10.1063/1.4928578} {\bibfield  {journal} {\bibinfo  {journal} {The Journal of
  Chemical Physics}\ }\textbf {\bibinfo {volume} {143}},\ \bibinfo {pages}
  {074106} (\bibinfo {year} {2015})}\BibitemShut {NoStop}%
\bibitem [{\citenamefont {Schild}\ \emph {et~al.}(2016)\citenamefont {Schild},
  \citenamefont {Agostini},\ and\ \citenamefont
  {Gross}}]{Schild16_Electronic_Flux}%
  \BibitemOpen
  \bibfield  {author} {\bibinfo {author} {\bibfnamefont {A.}~\bibnamefont
  {Schild}}, \bibinfo {author} {\bibfnamefont {F.}~\bibnamefont {Agostini}}, \
  and\ \bibinfo {author} {\bibfnamefont {E.~K.~U.}\ \bibnamefont {Gross}},\
  }\href {\doibase 10.1021/acs.jpca.5b12657} {\bibfield  {journal} {\bibinfo
  {journal} {The Journal of Physical Chemistry A}\ }\textbf {\bibinfo {volume}
  {120}},\ \bibinfo {pages} {3316} (\bibinfo {year} {2016})}\BibitemShut
  {NoStop}%
\bibitem [{\citenamefont {Eich}\ and\ \citenamefont
  {Agostini}(2016)}]{Agostini_adia_EF_16}%
  \BibitemOpen
  \bibfield  {author} {\bibinfo {author} {\bibfnamefont {F.~G.}\ \bibnamefont
  {Eich}}\ and\ \bibinfo {author} {\bibfnamefont {F.}~\bibnamefont
  {Agostini}},\ }\href {\doibase 10.1063/1.4959962} {\bibfield  {journal}
  {\bibinfo  {journal} {The Journal of Chemical Physics}\ }\textbf {\bibinfo
  {volume} {145}},\ \bibinfo {pages} {054110} (\bibinfo {year}
  {2016})}\BibitemShut {NoStop}%
\bibitem [{\citenamefont {Cohen}\ \emph {et~al.}(2025)\citenamefont {Cohen},
  \citenamefont {Steinitz-Eliyahu}, \citenamefont {Gross}, \citenamefont
  {Refaely-Abramson},\ and\ \citenamefont {Requist}}]{dmpv-zqdh}%
  \BibitemOpen
  \bibfield  {author} {\bibinfo {author} {\bibfnamefont {G.}~\bibnamefont
  {Cohen}}, \bibinfo {author} {\bibfnamefont {R.}~\bibnamefont
  {Steinitz-Eliyahu}}, \bibinfo {author} {\bibfnamefont {E.~K.~U.}\
  \bibnamefont {Gross}}, \bibinfo {author} {\bibfnamefont {S.}~\bibnamefont
  {Refaely-Abramson}}, \ and\ \bibinfo {author} {\bibfnamefont
  {R.}~\bibnamefont {Requist}},\ }\href {\doibase 10.1103/dmpv-zqdh} {\bibfield
   {journal} {\bibinfo  {journal} {Phys. Rev. B}\ }\textbf {\bibinfo {volume}
  {112}},\ \bibinfo {pages} {075102} (\bibinfo {year} {2025})}\BibitemShut
  {NoStop}%
\bibitem [{\citenamefont {Tu}\ and\ \citenamefont
  {Gross}(2025{\natexlab{a}})}]{nnkr-phm5}%
  \BibitemOpen
  \bibfield  {author} {\bibinfo {author} {\bibfnamefont {M.~W.-Y.}\
  \bibnamefont {Tu}}\ and\ \bibinfo {author} {\bibfnamefont {E.~K.~U.}\
  \bibnamefont {Gross}},\ }\href {\doibase 10.1103/nnkr-phm5} {\bibfield
  {journal} {\bibinfo  {journal} {Phys. Rev. Res.}\ }\textbf {\bibinfo {volume}
  {7}},\ \bibinfo {pages} {043075} (\bibinfo {year}
  {2025}{\natexlab{a}})}\BibitemShut {NoStop}%
\bibitem [{\citenamefont {Tu}\ and\ \citenamefont
  {Gross}(2025{\natexlab{b}})}]{tu2025nonadiabaticperturbationtheoryexact}%
  \BibitemOpen
  \bibfield  {author} {\bibinfo {author} {\bibfnamefont {M.~W.-Y.}\
  \bibnamefont {Tu}}\ and\ \bibinfo {author} {\bibfnamefont {E.~K.~U.}\
  \bibnamefont {Gross}},\ }\href {https://arxiv.org/abs/2511.02004} {\enquote
  {\bibinfo {title} {Non-adiabatic perturbation theory of the exact
  factorisation},}\ } (\bibinfo {year} {2025}{\natexlab{b}}),\ \Eprint
  {http://arxiv.org/abs/2511.02004} {arXiv:2511.02004 [physics.chem-ph]}
  \BibitemShut {NoStop}%
\bibitem [{\citenamefont {Li}\ \emph {et~al.}(2025)\citenamefont {Li},
  \citenamefont {Requist},\ and\ \citenamefont
  {Gross}}]{li2025bornoppenheimertimedependentdensityfunctional}%
  \BibitemOpen
  \bibfield  {author} {\bibinfo {author} {\bibfnamefont {C.}~\bibnamefont
  {Li}}, \bibinfo {author} {\bibfnamefont {R.}~\bibnamefont {Requist}}, \ and\
  \bibinfo {author} {\bibfnamefont {E.~K.~U.}\ \bibnamefont {Gross}},\ }\href
  {https://arxiv.org/abs/2511.09899} {\enquote {\bibinfo {title} {Beyond
  born-oppenheimer time-dependent density functional theory},}\ } (\bibinfo
  {year} {2025}),\ \Eprint {http://arxiv.org/abs/2511.09899} {arXiv:2511.09899
  [physics.chem-ph]} \BibitemShut {NoStop}%
\bibitem [{\citenamefont {Gonis}(2019)}]{GONIS20192772}%
  \BibitemOpen
  \bibfield  {author} {\bibinfo {author} {\bibfnamefont {A.}~\bibnamefont
  {Gonis}},\ }\href {\doibase https://doi.org/10.1016/j.physleta.2019.03.007}
  {\bibfield  {journal} {\bibinfo  {journal} {Physics Letters A}\ }\textbf
  {\bibinfo {volume} {383}},\ \bibinfo {pages} {2772} (\bibinfo {year}
  {2019})}\BibitemShut {NoStop}%
\bibitem [{\citenamefont {Wang}\ \emph {et~al.}(2025)\citenamefont {Wang},
  \citenamefont {Li},\ and\ \citenamefont {Li}}]{Wang_2025_Testing}%
  \BibitemOpen
  \bibfield  {author} {\bibinfo {author} {\bibfnamefont {Z.}~\bibnamefont
  {Wang}}, \bibinfo {author} {\bibfnamefont {Y.}~\bibnamefont {Li}}, \ and\
  \bibinfo {author} {\bibfnamefont {C.}~\bibnamefont {Li}},\ }\href {\doibase
  10.1063/5.0260749} {\bibfield  {journal} {\bibinfo  {journal} {The Journal of
  Chemical Physics}\ }\textbf {\bibinfo {volume} {162}},\ \bibinfo {pages}
  {234104} (\bibinfo {year} {2025})}\BibitemShut {NoStop}%
\bibitem [{\citenamefont {Scherrer}\ \emph {et~al.}(2017)\citenamefont
  {Scherrer}, \citenamefont {Agostini}, \citenamefont {Sebastiani},
  \citenamefont {Gross},\ and\ \citenamefont {Vuilleumier}}]{Rodolphe17PTmass}%
  \BibitemOpen
  \bibfield  {author} {\bibinfo {author} {\bibfnamefont {A.}~\bibnamefont
  {Scherrer}}, \bibinfo {author} {\bibfnamefont {F.}~\bibnamefont {Agostini}},
  \bibinfo {author} {\bibfnamefont {D.}~\bibnamefont {Sebastiani}}, \bibinfo
  {author} {\bibfnamefont {E.~K.~U.}\ \bibnamefont {Gross}}, \ and\ \bibinfo
  {author} {\bibfnamefont {R.}~\bibnamefont {Vuilleumier}},\ }\href {\doibase
  10.1103/PhysRevX.7.031035} {\bibfield  {journal} {\bibinfo  {journal} {Phys.
  Rev. X}\ }\textbf {\bibinfo {volume} {7}},\ \bibinfo {pages} {031035}
  (\bibinfo {year} {2017})}\BibitemShut {NoStop}%
\bibitem [{\citenamefont {Mátyus}\ and\ \citenamefont
  {Teufel}(2019)}]{Matyus19DBOC}%
  \BibitemOpen
  \bibfield  {author} {\bibinfo {author} {\bibfnamefont {E.}~\bibnamefont
  {Mátyus}}\ and\ \bibinfo {author} {\bibfnamefont {S.}~\bibnamefont
  {Teufel}},\ }\href {\doibase 10.1063/1.5097899} {\bibfield  {journal}
  {\bibinfo  {journal} {The Journal of Chemical Physics}\ }\textbf {\bibinfo
  {volume} {151}},\ \bibinfo {pages} {014113} (\bibinfo {year}
  {2019})}\BibitemShut {NoStop}%
\bibitem [{\citenamefont {Maskri}\ and\ \citenamefont
  {Joubert-Doriol}(2022)}]{Doriol22DBOC}%
  \BibitemOpen
  \bibfield  {author} {\bibinfo {author} {\bibfnamefont {R.}~\bibnamefont
  {Maskri}}\ and\ \bibinfo {author} {\bibfnamefont {L.}~\bibnamefont
  {Joubert-Doriol}},\ }\href {\doibase 10.1098/rsta.2020.0379} {\bibfield
  {journal} {\bibinfo  {journal} {Philosophical Transactions of the Royal
  Society A: Mathematical, Physical and Engineering Sciences}\ }\textbf
  {\bibinfo {volume} {380}},\ \bibinfo {pages} {20200379} (\bibinfo {year}
  {2022})}\BibitemShut {NoStop}%
\bibitem [{\citenamefont {Colbert}\ and\ \citenamefont
  {Miller}(1992)}]{DVR_Colbert_Miller92}%
  \BibitemOpen
  \bibfield  {author} {\bibinfo {author} {\bibfnamefont {D.~T.}\ \bibnamefont
  {Colbert}}\ and\ \bibinfo {author} {\bibfnamefont {W.~H.}\ \bibnamefont
  {Miller}},\ }\href {\doibase 10.1063/1.462100} {\bibfield  {journal}
  {\bibinfo  {journal} {The Journal of Chemical Physics}\ }\textbf {\bibinfo
  {volume} {96}},\ \bibinfo {pages} {1982} (\bibinfo {year}
  {1992})}\BibitemShut {NoStop}%
\bibitem [{\citenamefont {Surján}\ \emph {et~al.}(2024)\citenamefont
  {Surján}, \citenamefont {Ágnes Szabados},\ and\ \citenamefont
  {Gombás}}]{Surjan24_realeignH}%
  \BibitemOpen
  \bibfield  {author} {\bibinfo {author} {\bibfnamefont {P.~R.}\ \bibnamefont
  {Surján}}, \bibinfo {author} {\bibnamefont {Ágnes Szabados}}, \ and\
  \bibinfo {author} {\bibfnamefont {A.}~\bibnamefont {Gombás}},\ }\href
  {\doibase 10.1080/00268976.2023.2285034} {\bibfield  {journal} {\bibinfo
  {journal} {Molecular Physics}\ }\textbf {\bibinfo {volume} {122}},\ \bibinfo
  {pages} {e2285034} (\bibinfo {year} {2024})},\ \Eprint
  {http://arxiv.org/abs/https://doi.org/10.1080/00268976.2023.2285034}
  {https://doi.org/10.1080/00268976.2023.2285034} \BibitemShut {NoStop}%
\end{thebibliography}%


\end{document}